\documentclass[acmtog, authorversion]{acmart}

\usepackage{booktabs} 

\usepackage[ruled]{algorithm2e} 

\SetAlFnt{\small}
\SetAlCapFnt{\small}
\SetAlCapNameFnt{\small}
\SetAlCapHSkip{0pt}
\IncMargin{-\parindent}

\setcopyright{acmlicensed}
\acmJournal{TOG}
\acmYear{2017}\acmVolume{36}\acmNumber{4}\acmArticle{69}\acmMonth{7} \acmDOI{http://dx.doi.org/10.1145/3072959.3073643}

\usepackage{algorithm2e}
\usepackage{mathrsfs}
\usepackage{mathbbol}
\usepackage{mathtools}
\usepackage{overpic}
\usepackage{graphicx}
\usepackage{caption}
\usepackage{subcaption}
\usepackage[english]{babel}

\usepackage{color}
\definecolor{green}{rgb}{0.125,0.65,0.125}
\definecolor{reddy}{rgb}{0.7,0.,0.}
\definecolor{orange}{RGB}{255,74,13}
\definecolor{water}{rgb}{0.25,0.35,0.75}
\definecolor{pink}{rgb}{0.7,0.1,0.1}
\definecolor{highlightCol}{rgb}{0.1,0.72,0.25}

\newcommand{\addChanges}[1]{#1}
\newcommand{\addNChanges}[1]{#1}

\newcommand{\myrefeq}[1]{Eq.~(\ref{#1})}
\newcommand{\myreffig}[1]{Fig.~\ref{#1}}
\newcommand{\myreftab}[1]{Table~\ref{#1}}
\newcommand{\myrefsec}[1]{Section~\ref{#1}}

\newcommand{\myrefalg}[1]{Algorithm~\ref{#1}}

\newcommand{\vect}[1]{{\bf{#1}}}
\renewcommand{\v}[1]{{\bf{#1}}}

\newcommand{\desc}{\v{d}_w}
\newcommand{\ina}{\v{x}_1}
\newcommand{\inb}{\v{x}_2}
\newcommand{\nnw}{\v{w}}

\newcommand{\figsp}{0.0cm}
\setcitestyle{square}
\begin{document}
\title{Data-Driven Synthesis of Smoke Flows with CNN-based~Feature~Descriptors}
\author{Mengyu Chu}
\orcid{1234-5678-9012-3456}
\affiliation{
  \institution{Technical University of Munich}
}
\author{Nils Thuerey}
\affiliation{
	\institution{Technical University of Munich}
}

\renewcommand\shortauthors{Chu, M., Thuerey, N.}

\begin{abstract} 
We present a novel data-driven algorithm to synthesize high resolution flow simulations with reusable repositories of space-time flow data. In our work, we employ a descriptor learning approach to encode the similarity between fluid regions with differences in resolution and numerical viscosity. We use convolutional neural networks to generate the descriptors \addChanges{from fluid data such as smoke density and flow velocity}. At the same time, we present a deformation limiting patch advection method which allows us to robustly track deformable fluid regions. With the help of this patch advection, we generate stable space-time data sets from detailed fluids for our repositories. We can then use our learned descriptors to quickly localize a suitable data set when running a new simulation. This makes our approach very efficient, and resolution independent. We will demonstrate with several examples that our method yields volumes with very high effective resolutions, and non-dissipative small scale details \addChanges{that naturally integrate into the motions of the underlying flow}.

\end{abstract} 



\thanks{ This work is supported by the {\em ERC Starting Grant} 637014.  } 

\begin{teaserfigure}
	\centering
	\includegraphics[width=1.0\textwidth]{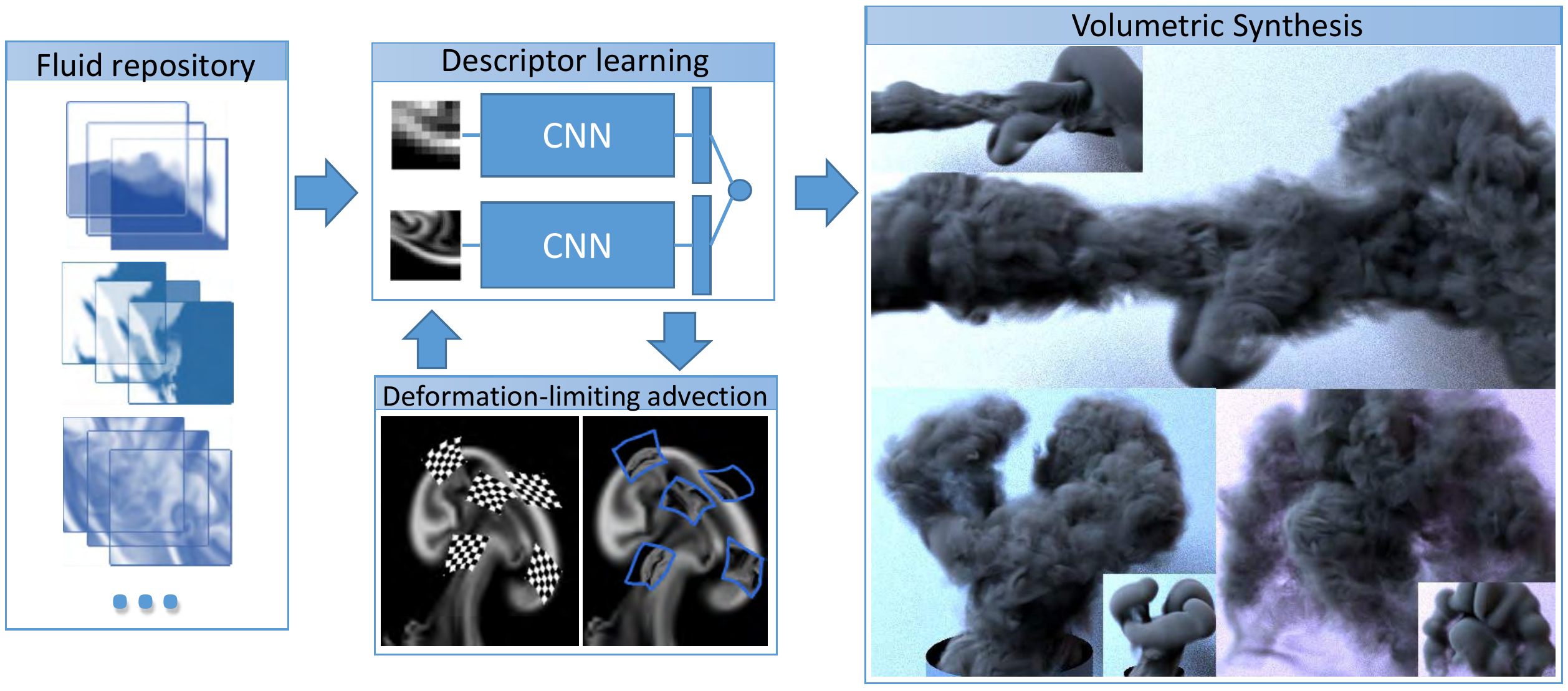}
	\caption{We enable volumetric fluid  synthesis with high resolutions and non-dissipative small scale details using CNNs and a fluid flow repository.}
	\label{fig:teaser} 
\end{teaserfigure}

\maketitle

\section{Introduction}
\label{sec:introduction}

Resolving the vast amount of detail of natural smoke clouds 
is a long standing challenge for fluid simulations in computer graphics.

Representing this detail typically requires very fine spatial resolutions,
which result in costly simulation runs, and in turn cause painfully long turn-around times.
A variety of powerful methods have been developed to alleviate this
fundamental problem: e.g., improving the accuracy of the advection step \cite{Kim05FlowFixer,Selle:2008:USM},
post-pro\-cessing animations with additional synthetic turbulence \cite{Kim:2008:wlt,narain:2008:procTurb},
and speeding up the pressure projection step \cite{lentine2010novel,ando2015dimension}.

We take a different perspective to efficiently realize high resolution flows: 
we propose to use a large collection of pre-computed space-time regions, from which
we synthesize new high-resolution volumes.
In order to very efficiently find the best match from this repository, we propose
to use novel, flow-aware feature descriptor.  We will ensure that $L_2$ distances in this feature
space will correspond to real matches of flow regions \addChanges{ in terms of fluid density as well as flow motion}, so that we can
very efficiently retrieve entries even for huge libraries of flow data sets.

Closely related to the goal of efficient high-resolution flows is the fight against numerical viscosity.
This is a very tough problem, as the discretization errors arising in practical flow settings
cannot be quantified with closed form expressions. 
On human scales the viscosity of water and air is close to zero, and the effects of numerical
viscosity quickly lead to unnaturally viscous motions for practical resolutions.
While we do not aim for a method that directly reduces
numerical viscosity, 
we will show that it is possible to predict
it's influence for typical smoke simulations in graphics. We do this in a data-driven fashion,
i.e., we train a model to establish correspondences between simulations with differing amounts
of numerical viscosity.

In our method, the calculation of descriptors and encoding the effects of discretization errors are 
handled by a convolutional neural network (CNN).
These networks were shown to be powerful tools 
to predict the similarity of image pairs \cite{mobahi2009deep,zagoruyko2015learning}, e.g., to compute
optical flow or depth estimates. We leverage the regressive capabilities of these CNNs to train
networks that learn to encode very small, yet expressive flow descriptors. These descriptors will encode  correspondences
in the face of numerical approximation errors, and at the same time allow for very efficient 
retrievals of suitable space-time data-sets from a repository.

We  compute these correspondences and repository look-ups for localized regions in the flow.
We will call these regions {\em patches} in the following,
and we employ a deforming Lagrangian frame to track the patches over time. 
Compared with recording data from static or rigid regions, this has the advantage that small features are pre-computed and stored in the repository, and do not inadvertently dissipate. In this way, we also side-step the strict
time step restrictions that fine spatial discretization usually impose. On the other hand, we have to make sure
the regions do not become too ill-shaped over time. For this, we propose a new deformation-limiting advection scheme with an anticipation step.
Motivated by the fractal nature of turbulence, and the pre-dominantly uni-directional energy transfer
towards small scales in Richardson's energy cascade \cite{Pope00}, we match and track each patch independently. 
This results in a very efficient method, as it allows us to perform all patch-based computations in parallel.

To summarize, the contributions of our work are: 
\begin{itemize}
	\item a novel CNN-based approach for computing robust, low-dimensional feature descriptors \addChanges{for density and velocity},
	\item a deformation limiting patch advection with anticipation, 
	\item an algorithm for
	efficient volumetric synthesis based on reusable space-time regions. 
\end{itemize}
In combination, our contributions make it possible
to very efficiently synthesize highly detailed flow volumes with the help of a re-usable space-time flow repository.
At the same time, we will provide an evaluation of convolutional neural networks and similarity learning
in the context of density-based flow data sets.

\section{Related Work}
\label{sec:relatedwork}

Fluid simulations for animation purposes typically leave out
the viscosity term of the {\em Navier-Stokes} (NS) equations, and solve the inviscid {\em Euler} equations: 
$ \text{D} \v{u} / \text{D} t  =-\nabla p + \v{g}$, where $\v{u},p$ and $\v{g}$ denote velocity,
pressure, and acceleration due to external forces, respectively.
These simulations have a long history in Graphics \cite{cg:kass:1990},
but especially for larger resolutions, computing a pressure field to make the flow divergence-free can become
a bottleneck. Different methods to speed up the necessary calculations have been proposed, e.g.,
multi-grid solvers \cite{McAdams:2010:PMP}, coarse projections \cite{lentine2010novel}, or lower-dimensional
approximations \cite{ando2015dimension}.
Another crucial part of Eu\-lerian solvers is computing the transport in the
grid. Here, unconditionally stable methods are widely used, especially the
semi-La\-grangian method \cite{stam1999} and its more accurate versions
\cite{Kim05FlowFixer,Selle:2008:USM}.
As we focus on single-phase flows, we restrict the following discussion to
corresponding works. For an overview of fluid simulations in computer
graphics we recommend the book by R. Bridson \cite{bridson2015book}.

While algorithms for reducing algorithmic complexity are vital for fast solvers, 
the choice of data-structures is likewise important.
Among others, recent works have proposed ways to handle
large-scale particle-based surfaces \cite{akinci2012parallel}, or highly efficient
tree structures \cite{museth2013vdb}.

Several works have also investigated viscosity for fluid animations, e.g., 
by proposing stable and accurate methods for discretization \cite{batty2008accurate},
efficient representations of viscous sheets \cite{batty2012discrete},
or illustrating the importance of viscosity for resolving shear flows near obstacles in the flow \cite{Zhang2016}.
While these are highly interesting extensions, most practical fluid solvers for computer animation omit solving for viscosity in order to reduce the computational work. 

\addChanges{Vortex-based methods aim for better preserving the swirling motions of flows,
which are typically highly important for detail and realism.
Methods to amplify existing vorticity have been proposed \cite{Fedkiw2001viscConf,selle2005vortex}, while other works model boundary layer effects \cite{Pfaff2009turbBoundary,zhu2010creating} or anisotropic vortices \cite{pfaff2010scalable}.
However, the amount of representable details is still inherently limited by the chosen grid resolutions for these algorithms. 
Hybrid methods have likewise been proposed to couple Eulerian simulations with Lagrangian vorticity simulations \cite{,Golas2012Decomposition}, while other researchers have proposed extrapolations of flow motions \cite{Zhang2013Extrapolation}. Our approach instead off-loads the computational burden to produce small-scale details to a pre-computation stage.}

So-called {\em up-res} techniques, which increase
the apparent resolution of a flow field, are another popular class of algorithms
to simulate detailed flows at moderate computational cost.
Different variations have been proposed for grid-based simulations 
\cite{Kim:2008:wlt,narain:2008:procTurb}, for mesh-based 
buoyancy effects \cite{pfaff2012lagrangian} or \addChanges{for particle-based simulation \cite{Mercier2015surfaceWave}}. 
While these works represent highly useful algorithms to generate
detailed animations, all small-scale features of a flow still have to be resolved and
advected at simulation time.  This causes significant amounts of computational
work. The complexity of the advection step with one of the semi-Lagrangian
methods is linear in the number of unknowns. 
Thus, decreasing the cell size from $\Delta x$ to $\Delta x / 2$ results in eight times more spatial degrees of freedom. In addition,
we typically have to reduce the time step size accordingly to prevent time integration errors from dominating the solution. 
This means there we face a roughly 16 times higher workload to compute the motion of advected quantities, such as the smoke densities.
In contrast, our method fully decouples the fine spatial scales  \addChanges{with the help of CNN-based descriptors. Thus our approach works with pre-computed
space-time regions of {\em real} flow data, instead of resorting to procedural models}. 
As such, we side-step the steep increase in complexity
resulting from very fine spatial discretization. Instead, our method purely works with low-resolution \addChanges{fluid data and low-dimension descriptors during simulation time. }
\addChanges{The only high resolution operation is the density synthesis, which can be 
performed on a per frame basis at render time, in a trivially parallel fashion. }

Our method shares similarities with previous approaches
for synthesizing textures for animated flows.
While early works focused on the problem of tracking texture data
on liquid surfaces \cite{bargteil2006semi,kwatra2007texturing},
a variety of interesting algorithms to synthesize two-dimensional fluid
textures has been developed over the years 
\cite{kwatra2005texture,narain2007feature}, with
various additions and improvements \cite{jamrivska2015lazyfluids}.
These texture synthesis approaches were extended to work with flow velocities
by Ma et al. \cite{ma2009motion}. Notably, this is the only
work performing the synthesis in three dimensions, as the cost 
for synthesizing 3D volumes is typically significant even for moderate sizes.
Unlike texture synthesis, we do not directly generate
a high-resolution output, but rather focus on efficiently and accurately
retrieving appropriate data sets from a large repository of pre-computed data.
In addition, our machine learning approach gives us the freedom
to encode both similarity and physical properties in the look-up. 

The {\em stamping} approach used in movie productions
similarly re-uses smaller regions of previous simulations \cite{wrennige2011}. 
However, this approach typically rigidly moves the stamps, and 
does not align their content with the simulation.
We found that a controllable deformation is a crucial component to make
the Lagrangian representation of the patches work. A similar idea
was explored previously for
animating two-dimensional flows with frequency controlled textures \cite{yu2010texadvect}.
While their algorithm uses a measure of the amount of deformation to blend in undeformed
content, we explicitly compute a new deformed state for our patches that takes into account
both the flow transport and undesirable deformations. This leads to increased life time
of the patch regions, and reduces blending operations correspondingly.

Machine learning with neural networks has been highly successful for a variety of challenging computer vision problems. In particular,
convolutional neural networks (CNNs) are particularly popular \cite{krizhevsky2012imagenet,simonyan2014very},
and several papers from this active area of research have targeted image similarity, e.g., to compute
depth maps \cite{zagoruyko2015learning}, or one-shot image classification \cite{koch2015cvpr}.
The first networks using a shared branch structure (so-called {\em Siamese}) were proposed 
by Bromley et al. \cite{bromley1993signature}, while learning descriptors with $L_2$ distances 
was employed for, e.g., descriptors of human faces \cite{chopra2005learning}.
In this context, we will employ a variant of the popular hinge loss function \cite{cao2006adapting,mobahi2009deep}, which represents 
the best convex approximation of a binary loss. 
While similarity of image pairs has been studied before, our aim is to work with 
density and velocity functions of fluid simulations. We will demonstrate
that it is beneficial to pay special attention to the velocity representation for learning.

Very few works so far exist that combine machine learning algorithms with animating fluids.
An exception is the regression-forest-based approach for smoothed particle hydrodynamics~\cite{ladicky2015data}. 
This algorithm computes accelerations for a Lagrangian NS solver based on \addChanges{carefully designed} input features,
with the goal to enable faster liquid simulations.
Our approach, on the other hand, aims for automatically extracting the best possible set of discriminative features.
We demonstrate that neural networks can perform this task very well,
and that we can in turn use the descriptors to establish correspondences between coarse and fine flow regions. 
Two other works share our goal to employ neural networks for 
single-phase flows: one focusing on learning the pressure projection step with a CNN \cite{tompson2016accelerating},
and another one learning local updates to remove divergence \cite{yang2016data}.
Especially the former of the two uses a deep convolutional network structure
that is similar to ours. 
However, their approach focuses on encoding the full
pressure field of a flow simulation with a neural network. In contrast, our networks learn 
robust descriptors for smaller regions of the flow. As such, our method can be easily
used with arbitrary resolutions. Both methods above work with the full output resolution, and
do not target the fine spatial resolutions we aim for with our method.

\section{Flow Similarity}
\label{sec:problem}

Given two flow simulations of the same effect,
one being a coarse approximation, the other one being a
more accurate version, e.g., based on a finer spatial discretization,
and a spatial region $\Omega$,
our goal is to compute a similarity score $s$ for the two inputs.
We consider functions $F_c$ and $F_f$ of the coarse and the fine flow, respectively, 
where $F$ could be a scalar value such as smoke density, or alternatively could also include the velocity, i.e.,
$F \in \mathbb{R}^3 \rightarrow \mathbb{R}^4$. 
We will  revisit which flow variables to include in $F$ in \myrefsec{sec:velocityDesc},
but for now we can assume without loss of generality that $F$ is a scalar function.
In order to compute similarity, we need to extract enough information from a region
of the flow such that $s$ can infer similarity from it. We will sample the flow functions
in a regular grid within $\Omega$, assuming that $F$ is sufficiently smooth to be
represented by point samples. All point samples from this grid are then combined
into an input vector $\v{x}_c$ and $\v{x}_f$ for coarse and fine simulations, respectively.

Given these inputs, we aim for computing $s(\v{x}_c,\v{x}_f)$ for $\Omega$ such that $s$ approaches zero if
the pair is actually one that corresponds to the same phenomenon
being represented on the coarse and fine scales.
For increasing dissimilarity of the flows, $s$ should increase. This dissimilarity
can, e.g., result from considering different regions $\Omega$ in the fine and coarse
simulations, or when the two are offset in time.

\begin{figure}[bt]
	\centering \includegraphics[width=0.45 \textwidth]{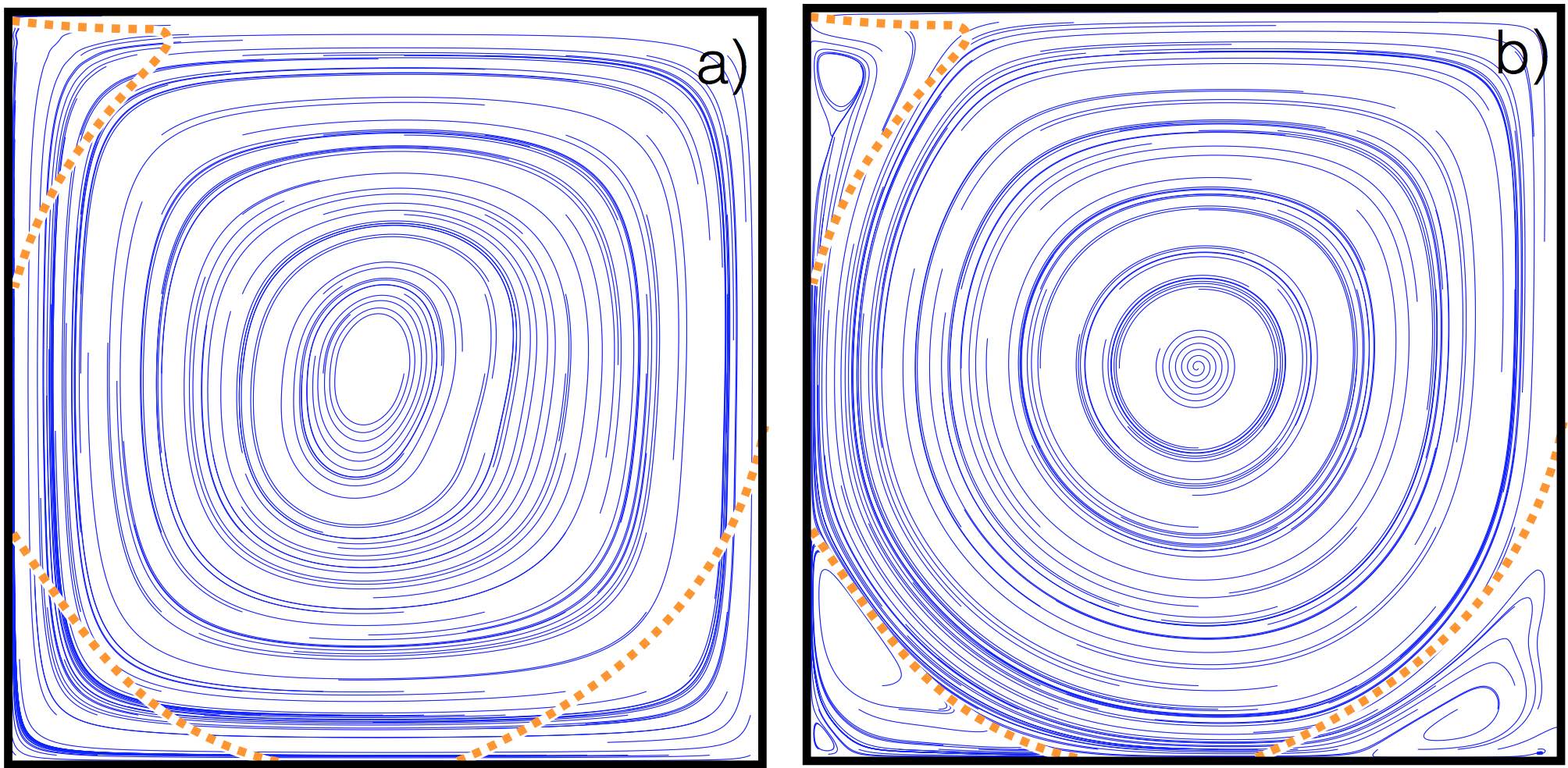}
	\caption{Stream lines of a lid driven cavity simulation without (left) and with viscosity (right) is shown. 
		The orange lines indicate the correct position of the center vortex \protect\cite{erturk2005numerical}. In this case, 
		the graphics approach commonly used to leave out viscosity leads to a very different vortex shape.}
	\label{fig:liddrivencavity} \vspace{\figsp}
\end{figure}

A first guess would be 
use an $L_2$ distance to compute $s$ as  $\int_{\Omega} \| \v{x}_f - \v{x}_c \|^2 d \v{x} $.
This turns out to be a sub-optimal choice, as 
even small translations can quickly lead to undesirably large distance values. 
Worse, in the presence of numerical viscosity, different resolutions for $F_c$ and $F_f$ can quickly lead to 
significantly different velocity and density content even when they should represent the same
fluid flow.
\myreffig{fig:liddrivencavity} illustrates how strongly viscosity can influence the outcome of a simulation. 
Numerical viscosity is typically comprised of errors throughout all parts of the
solver that influence velocity (although the advection step is
arguably the largest contributor). For practical algorithms, no closed form
solution exists to detect or quantify these errors.
Instead of manually trying to find heuristics or approximations of how
these numerical errors might propagate and influence solutions, we
transfer this task to a machine learning algorithm.

In addition, our goal is not only to measure the distance
between two inputs, but rather, given a new coarse input, we want to find
the best match from a large collection of pre-computed data sets.
Thus we map the correspondence problem to a feature space, such that distances
computed with a simple distance metric, e.g. Euclidean distance, corresponds to the desired similarity $s$. 
We will use a non-linear neural network to learn discriminative feature descriptors
$\v{d}(\v{x}) \in \mathbb{R}^m$, with $m$ as small as possible.
Given a coarse flow region $F_c$ we can then retrieve the best match
from a set of fine regions $F_f$ by minimizing $|| \v{d}(\v{x}_f) - \v{d}(\v{x}_c) ||^2$.

In the following, we will first describe our approach for
learning flow descriptors with CNNs. Afterwards we will explain our
deformation-aware patch tracking, which represents an important component
in order to achieve invariance with respect to large-scale motions.
Finally we explain how to synthesize high-resolution volumes for rendering.

\section{Learning Flow Similarity}
\label{sec:method}

As our approach focuses on computing distances between flow 
inputs with convolutional neural networks, we will briefly review the most important details below.
On a high level, we can view neural networks as a general methodology to approximate arbitrary
functions $f$, where we consider typical regression problems of the form $\v{y} = f( \v{x} , \v{w} )$,
i.e., given an input $\v{x}$ we want to approximate the result $\v{y}$ as closely as possible
given a representation $f$ based on the parameters $\v{w}$.
Thus, for our machine learning problems, the components of $\v{w}$ are the degrees of freedom we wish to compute.
In the following, we choose neural networks (NNs) to represent $f$,
for which a thorough overview can be found, e.g., in the book by
C. Bishop \cite{bishop2007pattern}.

NNs are represented by networks of nodes,
which, in contrast to our understanding of nature, are usually modeled with continuous functions.
A key ingredient are so-called activation functions $g$, that introduce non-linearity
and allow NNs to approximate arbitrary functions.
For a layer of nodes $L$ in the network, the output of the i'th node $y_{i,L}$ 
is computed with 
\begin{equation} \label{eq:nnLayer}
	y_{i,L} = g \Big( \sum_{j=0}^{n_{L-1}} w_{ij,L-1} \ y_{j,L-1} \Big) \ .
\end{equation}
Here, 
$n_L = $ denotes number of nodes in layer $L$, and, without loss of generality we assume $y_{0,L} = 1$.
This constant input 
for each node models a bias term, which 
is crucial to shift the output of the activation function. We employ this commonly used formulation to merge all degrees of
freedom in $\v{w}$. Thus, in the following, we will not distinguish between regular weights and biases.
We can rewrite \myrefeq{eq:nnLayer} using a weight matrix $W$ as
$\mathbf{y}_{L} = \mathbf{g} ( W_{L-1} \mathbf{y}_{L-1}) $.
In this equation, $\mathbf{g}$ is applied component-wisely to the input vector.
Note that without $\mathbf{g}$ we could fold a whole network with its multiple layers
into a single matrix $W_0$, i.e., $\mathbf{y} = W_0 \mathbf{x}$, which could
only be used to compute linearized approximations. Thus, in practice $\mathbf{g}$ is crucial to approximate generic, non-linear functions.
Typical choices for $\mathbf{g}$ are hyperbolic tangent or sigmoid functions.
For the learning process, the network is typically given a loss function $l( \v{y} , f( \v{x} , \v{w} ) )$,
that calculates the quality of the generated output with respect to $\v{y}$. The loss function needs to be differentiable
such that it's gradient can be back-propagated into the network, in order to update the weights.

A key component driving many of the recent deep-learning successes are so called {\em convolutional layers}.
These layers can be used to the exploit spatial structure of the input data, and essentially learn
filter kernels for convolutions. In the neural network area, these filters are often
called {\em feature maps}, as they effectively detect consistent features of the inputs that are important
to infer the desired output. This concept is especially useful for data such as sound (1D), or images (2D), and
directly extends to volumetric data sets such as those from fluid simulations.
The feature maps typically have only a small set of non-zero weights, e.g., 
$5\!\times\!5$ or $3\!\times\!3$ weights. As the inputs consist of vector quantities, e.g., RGB values for images, 
neural networks typically employ stacks of feature maps. Hence, we can think of the feature maps as higher-order
tensors. For volumetric inputs with three spatial dimensions storing a vector-valued function, we use
fourth-order tensors to represent a single feature map for a convolutional layer. Applying this feature map
yields a scalar 3D function. Most practical networks learn multiple feature maps at once per layer, and thus generate 
a vectorial value for the next layer, whose dimension equals the number of feature maps of the previous layer.

These convolutional layers are commonly combined with {\em pooling} layers which reduce the size of a
layer by applying a function, such as the maximum, over a small spatial region. Effectively, this down-samples
the spatially arranged output of a layer, in order to decrease the dimensionality of the problem for the next layer \cite{krizhevsky2012imagenet}.
Note that  convolutional neural networks in principle cannot do more than networks consisting only of fully connected layers.
However, the latter usually have a significantly larger number of  degrees of freedom. This makes them considerably more difficult 
to train, while convolutional networks lead to very reasonable network sizes and training times, making them very attractive
in practice. 
Smaller networks also have greatly reduced requirements for the amounts of input data, and can be beneficial for regularization \cite{bishop2007pattern}. 

Next we will explain the challenges in our setting when choosing a suitable loss function for the neural networks,
and outline the details of our network architectures afterwards.

\subsection{Loss Functions}

When computing our flow similarity metric, a first learning 
approach could be formulated as $s = f_s(\ina,\inb,\nnw)$, where $f_s$ is again the learned function represented by $\nnw$, $\ina$ and $\inb$
represent a pair of input features extracted from two simulations, 
and $s$ is the output indicating how similar the input pair is. 
As the inputs for our regression problems stem from a chaotic process, i.e. turbulent flow, the inputs look ``noisy'' from a regression
standpoint, and the training dataset is usually not linearly separable.
It is crucial that the learning process not only encodes a notion such as Euclidean distance of the two inputs, 
but also learns to use the whole space represented by its network structure 
to compute the similarity. At the same time, the network needs to reliably
detect dissimilar pairs. Hence, the availability of negative pairs is crucial for establishing robust similarity between true positives. 

The basic problem of learning similarity could be stated as follows:  given a pair of inputs, generate 
a label $y \in \{1,-1\}$, i.e., the pair is similar $(1)$, or not $(-1)$. 
While a naive  $L_2$ loss to learn exactly these labels
is clearly insufficient, a slightly improved loss function could be
formulated as $l_n= -y f(\ina,\inb, \nnw)$. In this case, the network would be
rewarded to push apart positive and negative pairs, but due to the
lack of any limit, the learned values would diverge to plus and minus infinity \cite{chapelle2006semi}.
Instead, it is crucial to have a  loss function that does not
unnecessarily constrain the regressor,
and at the same time gives it the freedom to push correctly classified pairs apart as much as necessary. 
The established loss function in this setting is the so called {\em hinge} loss, which can be computed with:
\begin{eqnarray} \label{eq:hingeLossBasic}
l_h(\ina,\inb) = \left\{ 
\begin{array}{l l}
\text{max}(0, 1 - f_s(\ina,\inb, \nnw)) \ ,  & y = 1 \\
\text{max}(0, 1 + f_s(\ina,\inb, \nnw)) \ , & y = -1
\end{array} \right.	
\end{eqnarray} 
This loss function typically leads to significant improvements over the naive loss functions outlined above.
When using a NN representation in conjunction with the loss function of \myrefeq{eq:hingeLossBasic}, a feature descriptor 
can be extracted by using the outputs of the last fully connected layer as a descriptor~\cite{zagoruyko2015learning}. 

While this approach works, we will demonstrate that it is even better to embed the $L_2$ distance
of the descriptors directly into the hinge loss \cite{mobahi2009deep}. 
As we later on base our repository lookups on these distances,
it is important to guide the NN towards encoding discriminative distances based on the feature descriptors themselves,
instead of only optimizing for a final similarity score $s$.
In order to do this, we can re-formulate the learning problem to generate the descriptor itself, using the descriptor 
distance as the "dissimilarity". In the following we will denote the outputs
of a specific {\em descriptor} layer of our network with $\desc(\v{x})$, where $\v{x}$ is the
input for which to compute the descriptor.
Based on these descriptors we change the regression problem to 
$ f_e(\ina,\inb, \nnw) = \beta - \alpha || \desc(\ina) - \desc(\inb) ||, \alpha>0 $.
Here we have introduced the parameters $\alpha$ and $\beta$ to fine
tune the onset and steepness of the function.
Using $f_e$ instead of $f_s$ in \myrefeq{eq:hingeLossBasic} yields 
\begin{eqnarray} \label{eq:hingeEmbedded}
\resizebox{.92\hsize}{!}{$
	l_e(\ina,\inb) = \left\{ 
	\begin{array}{l l}
	\text{max}(0, \alpha_p + ||\desc(\ina) - \desc(\inb)||  ) \ , & y = 1 \\
	\text{max}(0, \alpha_n - ||\desc(\ina) - \desc(\inb)||  ) \ ,  & y = -1 \ ,
	\end{array} \right.
	$}
\end{eqnarray} 
where we have replaced the $\alpha, \beta$ parameters by $\alpha_{p,n}$ which can
be used to fine tune the margins individually for positive and negative pairs, as we will discuss below.
Note that we normalize the descriptors $\desc$ when extracting them for \myrefeq{eq:hingeEmbedded}. 
This significantly improves convergence, and supports learning distributions of components,
rather than absolute values in the descriptor.
We will demonstrate that this loss function clearly outperforms the other alternatives,
after describing the details of our neural networks to be used in conjunction with this loss function.

\begin{figure}[t]
	\centering \includegraphics[width=0.45 \textwidth]{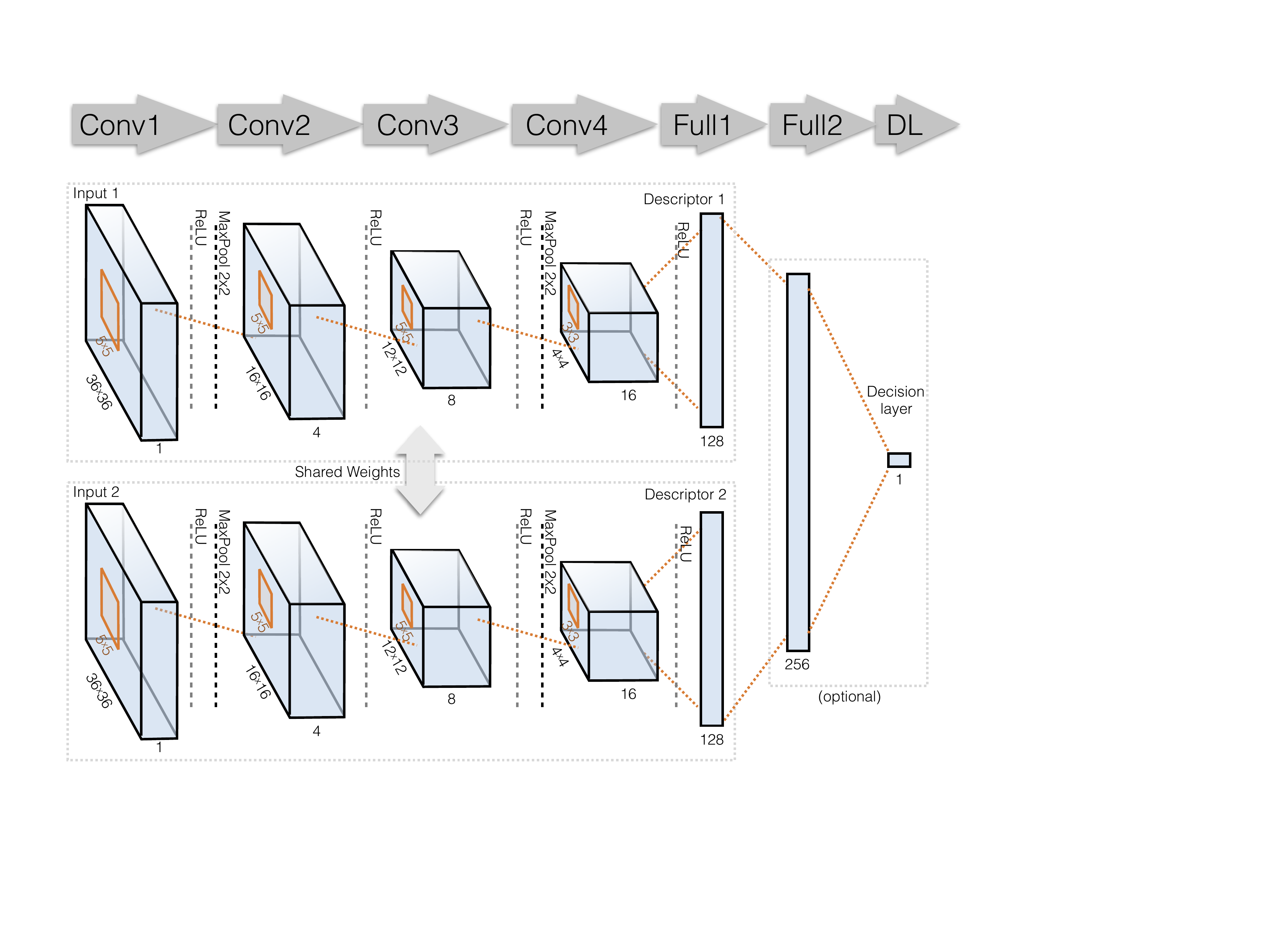}
	\caption{ Our CNN architecture with two convolution stacks with shared weights,
		followed by a feature and an optional decision layer. } 
	\label{fig:network} \vspace{\figsp}
\end{figure}

\subsection{CNN Architecture}
\label{sec:NNarchitecture}

Our neural networks consist of a typical stack of convolution layers which translate spatially arranged data
into an increasing number of feature signals with lower spatial resolution. A visual summary of the network
is given in  \myreffig{fig:network}. As our network compares two inputs, it initially
has two branches which contain a duplicated  stack of convolutional layers
with shared weights. Each branch acts separately on one input vector to reduce its dimensionality.
The outputs of the last convolutional layer of each branch ({\em Conv3} in our case) are concatenated into a 
serial vector containing the two feature descriptors (layer {\em Full1}).
For networks that regress a single similarity score with loss function like \myrefeq{eq:hingeLossBasic},
we add another fully connected layer ({\em Full2}) and an output layer with a single output node that computes the final similarity score. However, these two layers are optional.
When training with the hinge embedding loss, \myrefeq{eq:hingeEmbedded}, we omit these two layers.

We will use the following abbreviations to specify the network structure:
convolutional layer (CL), max pooling layer (MP), and fully connected layer (FC).
We start with inputs of size 36x36 in 2D. The input from a low-resolution
simulation is linearly up-scaled to this resolution (typically we use a four times
lower resolution for the coarse simulation).
One convolutional branch of our network 
yields $2^2$ points with 32 features each, i.e., 128 values in total. 
These 128 outputs are concatenated into the final feature descriptor layer  $\v{d}_w$ with normalization. For three dimensional inputs, we extend the spatial dimension in each layer correspondingly.
Hence, the first layer has a resolution of 5x5x5x4 in 3D, and the final spatial resolution is $2^3$ with 32 features.
Thus in 3D, our feature descriptor $D_w$ has dimension 256.

\subsection{Data Generation}
\label{sec:TrainingDataGen}

For machine learning approaches it is crucial to have good training data sets. Specifically, 
in our setting the challenge is to create a controlled environment for simulations
with differing discretizations, and resulting in different numerical viscosities.
Without special care, the coarse and fine versions will deviate over time, and due to their
non-linearity, initial small differences will quickly lead to completely different flows.
In the following, we consider flow similarity within a chosen time horizon $t_r$.
Note that this parameter depends on the setup under consideration, e.g., for very smooth
and slow motions, larger values are suitable, while violent and fast motions mean flows
diverge faster. \addChanges{We will discuss the implications of choosing $t_r$ in more detail below.}

We use randomized initial condition to create our training data. Given an initial condition,
we set up two parallel simulations, one with a
coarse resolution of $r_c$ cells per axis, and we typically use a four times higher resolution \addChanges{$r_f = 4r_c$} for the fine version.
While it would be possible to simply run a large number of simulations for a time $t_r$, we found that it is
preferable to instead synchronize the simulation in intervals of length $t_r$. Here, we give priority
to the high resolution, assuming that it's lower numerical viscosity is closer to the true solution.
We thus re-initialize the coarse simulation in intervals of length $t_r$ \addChanges{with a low-pass filtered version} of the fine simulation.

This synchronization leads to a variety
of interesting and diverse flow configurations over time, which we would otherwise have to recreate
manually with different initial conditions. For our data generation, we found buoyant flows to be problematic in rectangular domains due to their rising motion.
Using tall domains would of course work, but typically wastes a significant amount of space. Instead,
we compute a center of mass for the smoke densities during each time step. We then add a correction vector
during the semi-Lagrange advection for all quantities to relocate the center of mass to the grid center. During the data generation runs, we seed patches throughout the volume, and track them
with the same algorithm we use for synthesis later on. Thus, we also employ our deformation limiting there, which we 
describe in detail in \myrefsec{sec:defolim}. 
For each patch region, we record the full coarse and fine velocity / density functions 
within each deforming patch region for each time step. Currently, we track the patches 
with the fine simulation, and use the same spatial region in the coarse simulation. 

The recorded pairs of spatial regions for one time step give us the set of positive pairs for training. 
Note that coarse and fine data in these regions may have diverged up to the duration $t_r$.
To create negative pairs, we assign a random fine data set to each coarse input.
Two features from a negative pair are either recorded by different patches, or recorded by the same patch, but in different time steps. Therefore, for any coarse feature example $\ina$ in our training and evaluation dataset, there is only one fine feature example $\inb$ marked as relevant. 

\begin{figure}[tbh]
	\centering  
	\includegraphics[width=0.45 \textwidth]{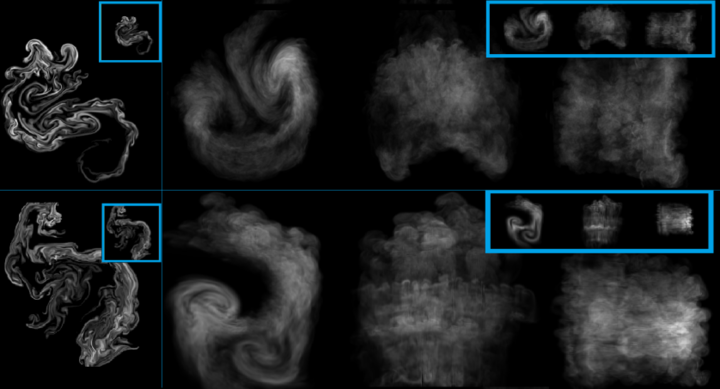}
	\caption{ Examples of our data generation for training, in both 2D (left) and 3D (front, left and top views). 
		The coarse simulation (with blue outline) is synchronized with the high resolution data in intervals $t_r$. }
	\label{fig:datagen} \vspace{\figsp}
\end{figure}

In this way, we have created several combined simulations in both 2D and 3D, with $t_r = 20$ and $t_r = 40$, to generate training datasets 
as well as evaluation datasets. \addChanges{These $t_r$ are selected so that the resulting training data have a maximum discriminative capabilities. For smaller intervals, the network presumably only sees very similar inputs, and hence cannot generate expressive descriptors. When the interval becomes too large, inputs can become too dissimilar for the NN. In general, $t_r$ is negatively correlated to the time step, kinetic energy and resolution difference. We currently select the $t_r$ manually through comparisons, and an automatic calculation of $t_r$ remains an interesting future direction.}

Several images of our data generation in 2D and 3D can be found in \myreffig{fig:datagen}. The detailed simulations have a 4 times higher resolution. For training, we generated 18,449 positive pairs for 2D and 16,033 pairs in total for 3D. 
In 2D, every training batch contains 1:1 positive and negative pairs. The latter ones are randomly generated from all positive ones while training. 
\addChanges{While a ratio of 1:1 was sufficient in 2D, training with this ratio turned out to be slow in 3D.
We found that increasing the number of negative pairs improves the learning efficiency, while having negligible influence on the converged state of the CNNs. Thus, we use a ratio of 1:7 for 3D training runs.}
For our evaluations below, the datasets with $t_r = 20$ and $t_r = 40$ give very similar results. Since the latter one shows a clearer differences between the methods,
in \myrefsec{sec:eval}, we will focus our evaluations on the dataset with $t_r = 40$ , which has 5440 and 5449 positive pairs in 2D and 3D respectively.

\subsection{Evaluation}
\label{sec:eval}

In order to evaluate and compare the success of the different approaches
it would be straight forward to compute descriptors with a chosen method for
a coarse input $i$, and then find the best match from a large collection of fine pairs.
If the best match is the one originally corresponding to $i$, we can count this as a success,
and failure otherwise. In this way, we can easily compute a percentage of successfully
retrieved pairs. 
However, this metric would not represent
our application setting well. Our goal is to employ the descriptors for patches in new simulations,
that don't have a perfectly corresponding one in the repository. For these we want to robustly retrieve the closest
available match. 
Thus, rather than counting the perfect matches, we want to evaluate how reliably our networks
have learned to encode the similarity of the flow in the descriptor space. To quantify this reliability,
we will in the following measure the true positive rate, which is typically called {\em recall}, over the cut-off rank {\em k}.

The recall over a cut-off rank is commonly employed in the information retrieval field to evaluate ranked retrieval results~\cite{Manning2008IIR}. 
Recall stands for the percentage of correctly retrieved data sets over all given related ones. The rank in this
case indicates the number of nearest neighbors that are retrieved from the repository for a given input.
In particular, for our evaluation dataset with $N$ pairs, with a given cut-off $k$, we evaluate the recall for all $N$ coarse features, and thus $kN$ pairs are retrieved in total per evaluation. In these retrieved pairs, if $r$ pairs are correctly labeled as related, the recall at cut-off $k$ would be $r/N$.
In such a case, a perfect method, would yield 100\% for the recall at $k=1$, and then be constant for larger $k$. In practice, methods will
slowly approach 100\% for increasing $k$, and even the worst methods will achieve a recall
of 100\% for $k = N$. Thus, especially a first range of small $k$ values
is interesting to evaluate how reliably a method has managed to group similar data in the Euclidean
space of the descriptor values.

\begin{figure}[tbh]
	\centering 
	\includegraphics[width=0.45 \textwidth]{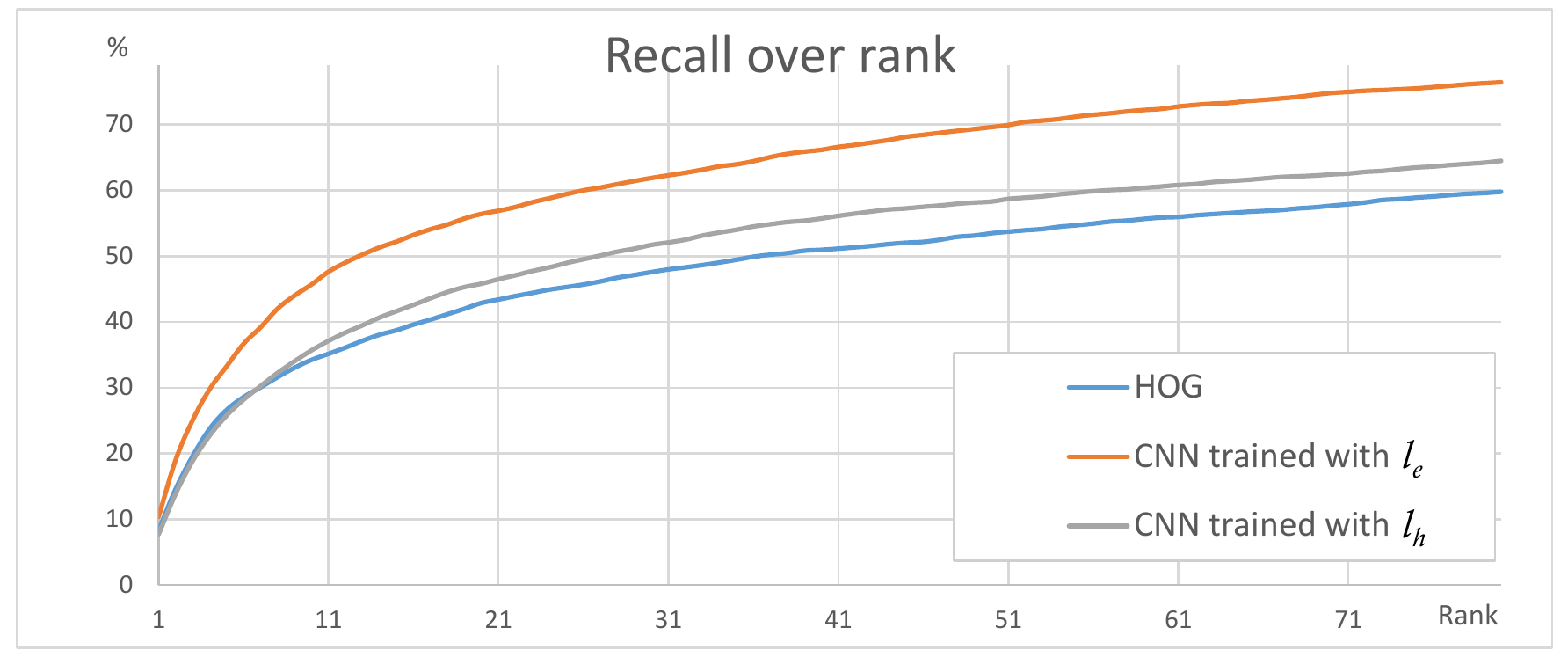}
	\caption{Recall over rank for HOG (blue), CNN trained with $l_h$ on similarity output node (gray), and CNN trained with $l_e$ on descriptors directly (orange). }
	\label{fig:RSgeneral} \vspace{\figsp}
\end{figure}

We first compare two CNN-based descriptors created with the two loss functions explained above,
and the popular, hand-crafted HOG descriptor in 2D. The latter is a commonly employed, and very successful feature
descriptor based on histograms of gradients of a local region.
As can be seen in \myreffig{fig:RSgeneral}, the HOG descriptor fares worst in this setting. Beyond a rank of 6, it's 
recall is clearly below that of the regular hinge loss CNN descriptor. The hinge embedded loss function
yields the best descriptor, which in this case is consistently more than 10\% better from rank 10 on.
The high recall rates show that our CNN successfully learns to extract discriminative features,
and can do so with a higher accuracy than conventional descriptors.

We also investigate the influence of the threshold $\alpha_p$ and $\alpha_n$ in the loss function $l_e$ in \myrefeq{eq:hingeEmbedded}.
As the parametrization $ [0.0, 0.7]$ has a slightly higher accuracy among others, 
we will use these parameters in the following. 
We found that it is not necessary to have any margin on the positive side of the loss function $l_e$, 
but on the negative side, a relatively large margin gives our CNN the ability to better learn the dissimilarities of pairs.

\begin{figure}[tbh]
	\begin{subfigure}[b]{0.21\textwidth}
		\includegraphics[width=\linewidth]{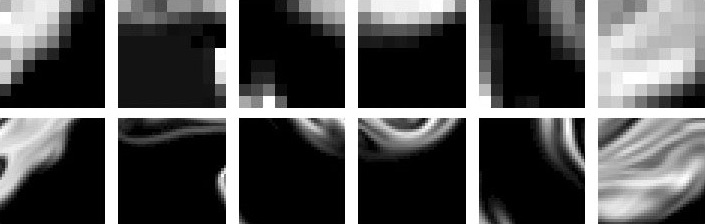}
		\caption{True Positive}
		\label{fig:TP}
	\end{subfigure}%
	\quad
	\begin{subfigure}[b]{0.21\textwidth}
		\includegraphics[width=\linewidth]{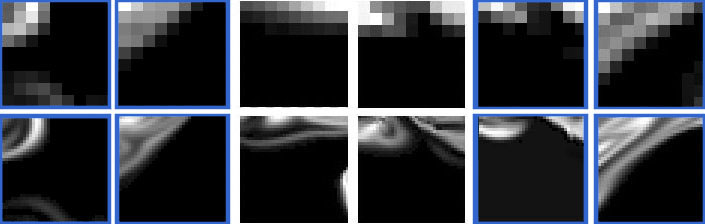}
		\caption{False Positive}
		\label{fig:FP}
	\end{subfigure}%
	\\
	\begin{subfigure}[b]{0.21\textwidth}
		\centering
		\includegraphics[width=\linewidth]{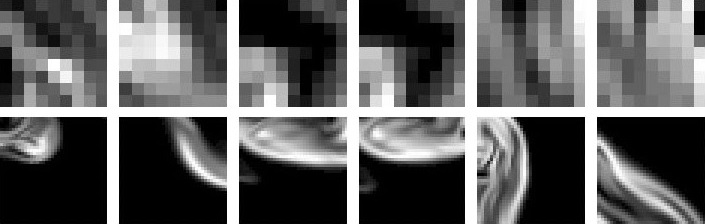}
		\caption{True Negative}
		\label{fig:TN}
	\end{subfigure}%
	\quad
	\begin{subfigure}[b]{0.21\textwidth}
		\centering
		\includegraphics[width=\linewidth]{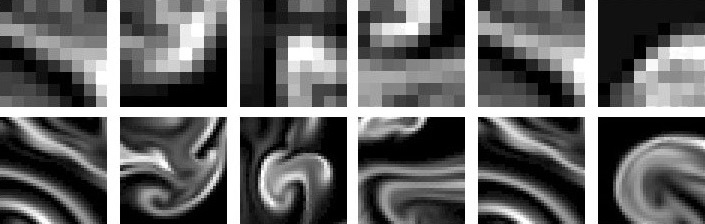}
		\caption{False Negative}
		\label{fig:FN}
	\end{subfigure}
	\caption{Top-ranking density pairs matched by our CNN.}\label{fig:evalExample}
\end{figure}

\myreffig{fig:evalExample} shows some of the top ranking true and false correspondences made by our CNN for the 
smoke density pairs. Correct positive and negative pairs are shown  on the left.
False negative pairs are related ones, for which the CNN descriptors still have a large difference, while the false positive ones 
are mistakenly matched pairs which were not related. 
In practice, the false negative pair have no effects on the synthesized results, in contrast to the false positives.
However, we notice that these false positives typically contain visually very similar data. As such,
these data sets will be unlikely to introduce visual artifacts in the final volume. Some of these false positives
actually stem from the same tracked patch region during data generation, and were only classified wrongly
in terms of their matched time distance. These pairs are marked with blue borders in \myreffig{fig:evalExample}(b).

\subsection{Descriptors for Flow Motions}
\label{sec:velocityDesc}

Up to now we have only considered smoke densities for establishing correspondences,
however, in the fluid simulation context we also have velocity information. The velocities
strongly determine the smoke motion over time, and as such they are likewise important
for making correspondences between the data of a new simulation and the data-sets in the repository.

To arrive at a method that also takes the flow motion into account, we use
two networks: one is trained specifically for density descriptors, while
we train the second one specifically for the motion. This yields two descriptors,  $\v{d}_{den}$ and $\v{d}_{mot}$,
which we concatenate into a single descriptor vector for our repository lookups with
\begin{equation} \label{eq:DescipCombine}
\v{d}(\v{x}) = \frac{1}{\sqrt{1+w_m^2}}{\begin{bmatrix}\v{d}_{den}(\v{x})\\[1mm]
	w_{m}\v{d}_{mot}(\v{x})\end{bmatrix}} \ .
\end{equation}
Note that the separate calculation of density and motion descriptor mean
that we can easily re-scale the two halves to put more emphasis on one
or the other. 
Especially when synthesizing new simulations results, we put more emphasis on the density content with  $w_{m} = 0.6$.

\begin{figure}[bht]
	\centering 
	\includegraphics[width=0.45 \textwidth]{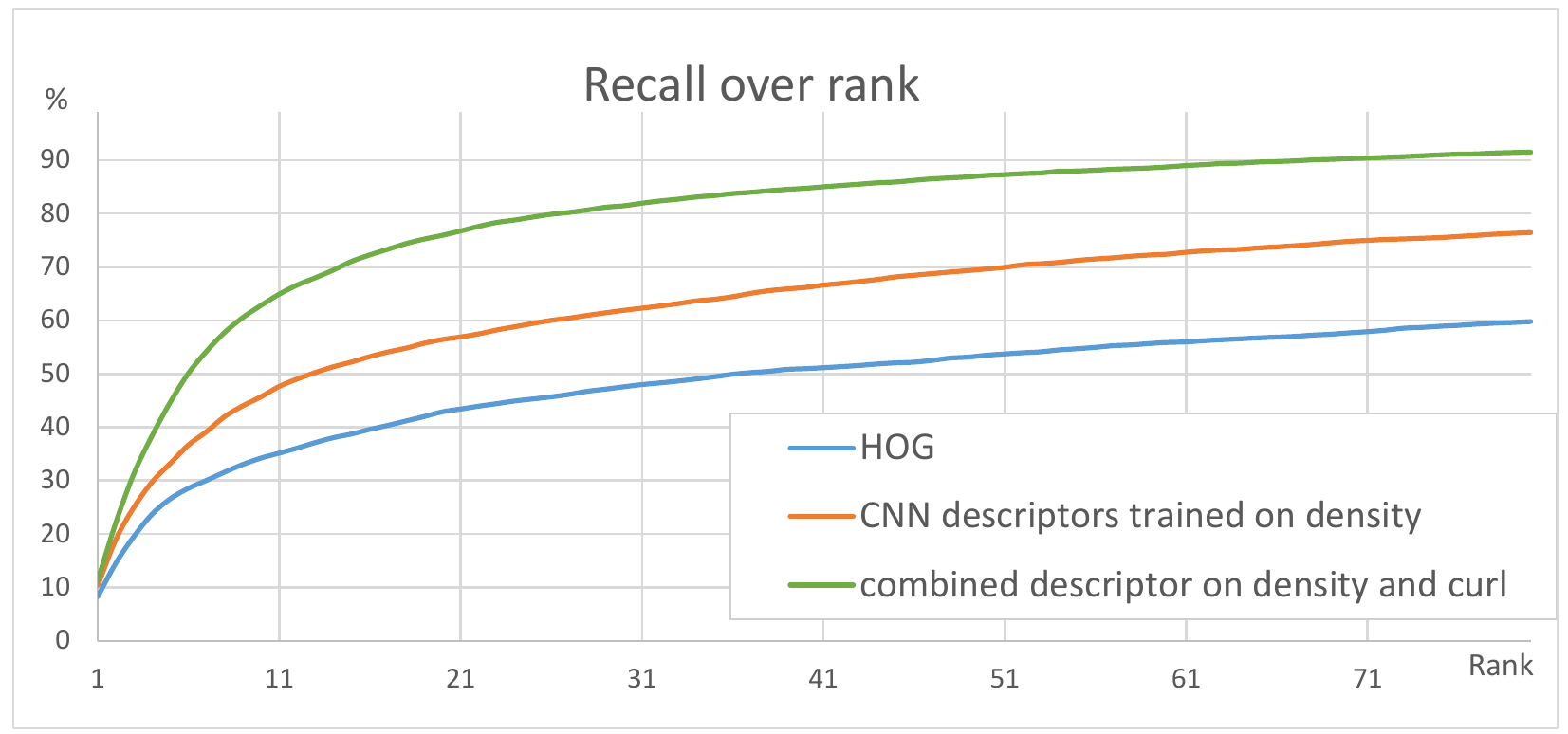}
	\caption{Using curl descriptors as well as density descriptors improves matching performance. }
	\label{fig:RSoutput} \vspace{\figsp}
\end{figure}

\begin{figure}[bht]
	\centering 
	\includegraphics[width=0.45 \textwidth]{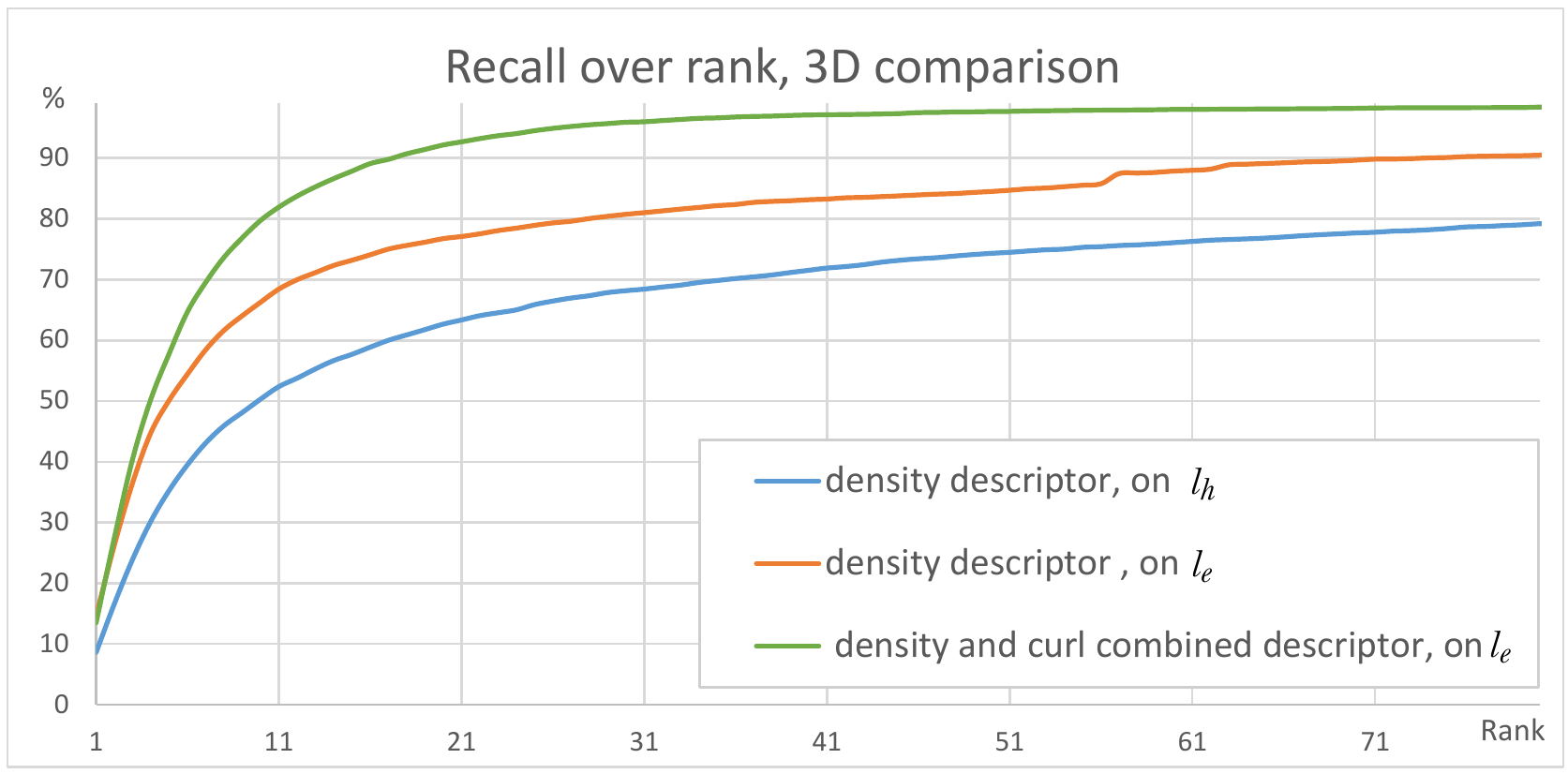}
	\caption{In 3D, using curl descriptors as well as density descriptors improves matching performance. }
	\label{fig:RSoutput3D} \vspace{\figsp}
\end{figure}

For the motion descriptor, we use the vorticity as input, i.e., 
$\omega = \nabla \times \v{u}$. 
\addChanges{During the synthesis step, motion descriptors generated from vorticity 
offer significantly better look-ups than the ones trained with $\v{u}$,} as the vorticity better reflects local changes
in the flow field. Due to our scale separation with patch motion and content, our
goal is to represent local, relative motions with our descriptors, instead of, e.g.,
large scale translations or stretching effects. 
Using a combined density and curl descriptor improves the recall rate even further.
A comparison with our 2D data set is shown in \myreffig{fig:RSoutput}, e.g., at rank 11, the
recall improves by ca. 35\%, and we see similar gains in three dimensions
(shown in \myreffig{fig:RSoutput3D}). Note that these two figures use
a weight of $w_{m} = 1.0 $ for the curl descriptor.

Due to the aforementioned improvements in matching accuracy, this approach represents 
our final method. In the following we will use a double network, one trained for densities
and a second one trained for the curl to compute our descriptors.

\subsection{Direct Density Synthesis}

Seeing the generative capabilities of modern neural networks, we found it
interesting to explore how far these networks could be pushed in the context of
high-resolution flows. Instead of aiming for the calculation of a
low-dimensional descriptor, the NN can also be given the task to directly regress a high
resolution patch of smoke densities.  An established network structure for this goal 
is a stack of convolutional layers to reduce the input region to a
smaller set of feature response functions, which then drive the generation of
the output with stack of convolution-transpose layers \cite{krizhevsky2012imagenet}. 

\begin{figure}[bt]
	\centering 
	\includegraphics[width=0.47 \textwidth]{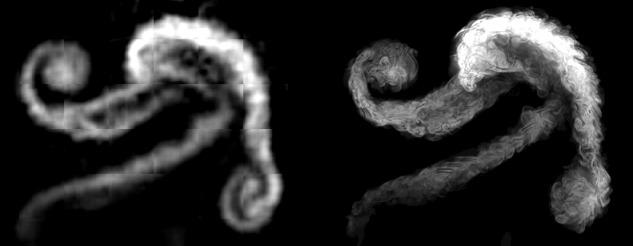}
	\caption{ Directly trying to synthesize densities with CNNs yields
		blurred results that lack structures (left). Our algorithm computes
		highly detailed flows for the same input (right).
	}
	\label{fig:hook} \vspace{\figsp}
\end{figure}

We ran an extensive series of tests, and the best results we could achieve
for a direct density synthesis are shown in \myreffig{fig:hook}.
In this case, the network receives a region of 16x16 density values,
and produces outputs of four times higher resolution (64x64) with the help of two convolutional
layers, a fully connected layer, and four deconvolutional layers.
While we could ensure convergence of the networks without overfitting,
and relatively good temporal stability, 
the synthesized densities lack any detailed structures.
This lack of detail arises despite the fact that this network has a significantly
larger number of weights than our descriptor network, and had more training
data at its disposal.

There is clearly no proof that it is impossible to synthesize detailed
smoke densities with generative neural networks, but we believe that our tests
illustrate that the chaotic details of turbulent smoke flows
represent an extremely challenging task for NNs. Especially when trying to
avoid overfitting with a sufficiently large number of inputs,
the turbulent motions seem like noise to the networks.
As a consequence, they learn an {\em averaged} behavior that smoothes out detailed structures.
These results also motivate our approach: we side-step these problems by
learning to encode the distance between flow regions, and supplying best matched details in our flow repository at render time, instead of learning and generating details directly.

\section{Motion and Synthesis}
\label{sec:motion}

For each patch, we track its motion with a local grid. 
We call these grids 
{\em cages} in the following, to distinguish them from the Cartesian grids of the fluid simulations,
and we will denote the number of subdivision per spatial axis with $n_{\text{cage}}$,
with a resulting cell size $\Delta x_{\text{cage}}$.
Below, we describe our approach to limit excessive deformation of these cages.

\subsection{Deformation-limiting Motion}
\label{sec:defolim}

\begin{figure}[t]
	\includegraphics[width=0.47 \textwidth]{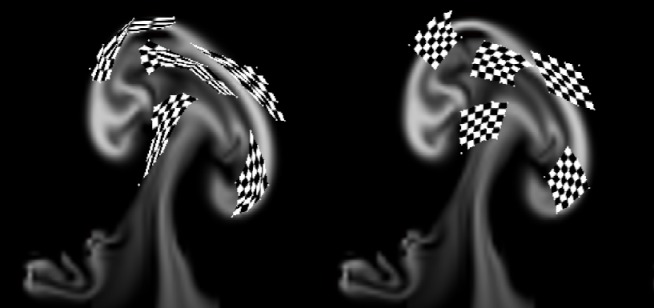}
	\caption{ 
		Naive patch advection (left) quickly leads to distorted regions. 
		Our deformation limiting advection (right) can keep stable regions while following the fluid motion.
	}
	\label{fig:patchCompare} \vspace{\figsp}
\end{figure}

Even simple flows exhibit large amounts of rotational motion, that quickly lead to very undesirable 
stretching and inversion artifacts. 
An example can be seen on the left side of \myreffig{fig:patchCompare}. 
Thus, when advecting the Lagrangian patch regions through a velocity field,
we are facing the challenge to avoid overly strong distortions while making
the patch follow the pre-scribed motion. 
Inspired by previous work on {\em as-rigid-as-possible} deformations \cite{sorkine2004,igarashi2005}, we express
our Lagrangian cages in terms of differential
coordinates, and minimize an energy functional in the least-squares sense
to retrieve a configuration that limits deformation while adhering to the flow motion.

\newcommand{\cEdge}[2]{\v{e}_{{#1},{#2}}}
\newcommand{\cRot}[2]{R_{{#1},{#2}}}

\begin{figure}[tbh]
	\centering 
	\includegraphics[width=0.3 \textwidth]{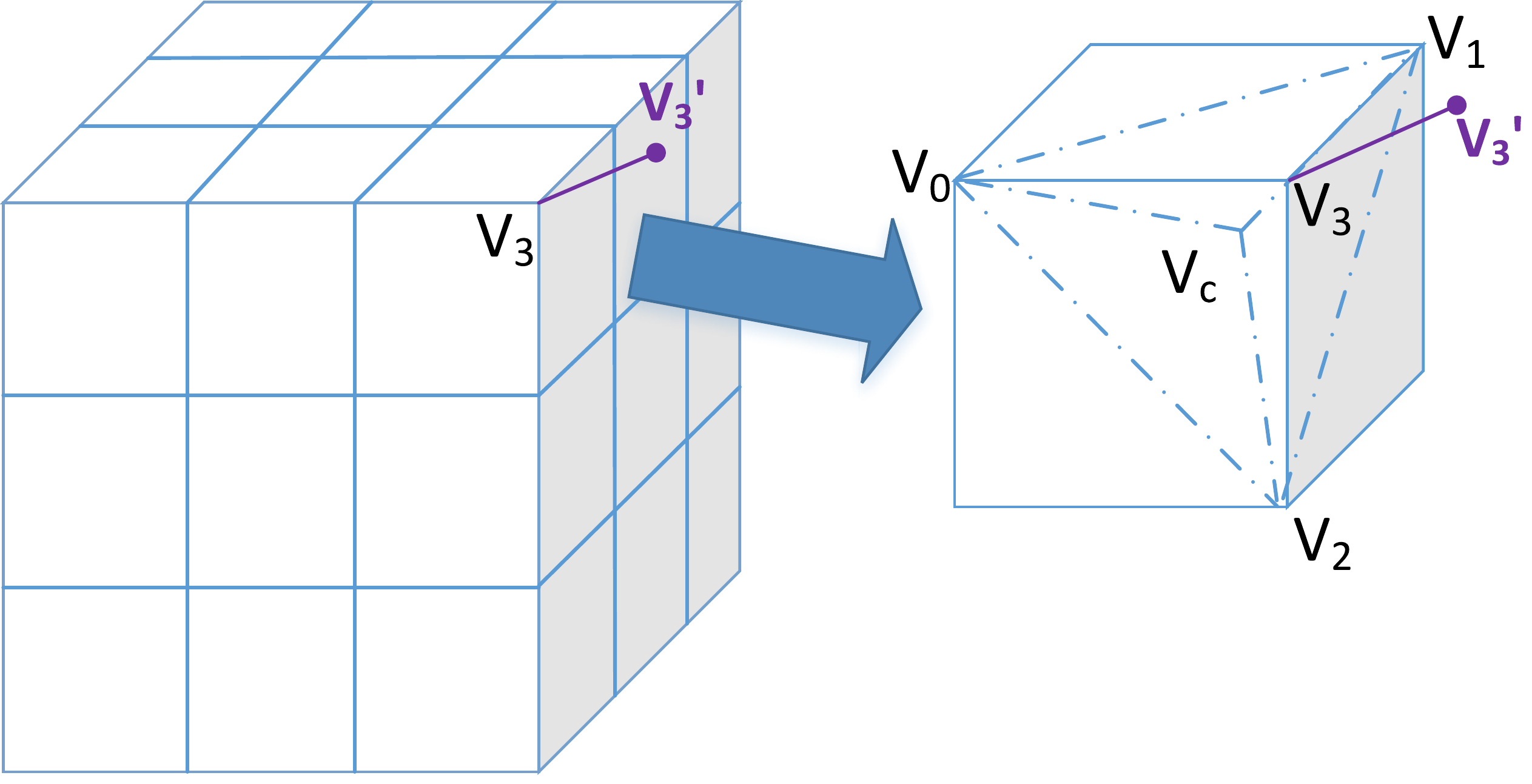}
	\caption{ An example of patch cage with $n=3$. The deformation error 
		is accumulated from every cell, as shown on the right. The distance between 
		a target position $v_3$ and the advected position $v_3'$ yields the 
		deformation error. 
		$v_3$ is expressed in terms of $v_{0,1,2}$ and their center point $v_c$. 
	}
	\label{fig:cube} \vspace{\figsp}
\end{figure}

For the differential coordinates, we span an imaginary tetrahedron between
an arbitrary vertex $\v{v}_3$, and its three neighboring vertices $\v{v}_{0,1,2}$, as shown in \myreffig{fig:cube}.
For the undeformed state of a cell, the position of $\v{v}_3$ can be expressed with rotations
of the tetrahedron edges as 
\begin{equation} \label{eq:v3desire}
\v{v}_3 = \v{v}_c +
	{(R_{\cEdge{1}{0}}\cEdge{p}{2}+
	R_{\cEdge{2}{1}}\cEdge{p}{0}+ 
	R_{\cEdge{0}{2}}\cEdge{p}{1})}/ {3\sqrt{2}} \ .
\end{equation}
Here $\cEdge{i}{j}$ denotes the edge between points $i$ and $j$, and
$R_\v{v}$ is the the 3x3 rotation matrix that rotates by $90$ degrees around axis $\v{v}$.
$\v{v}_c$ is the geometric center of the triangle spanned the three neighbors, i.e., 
$\v{v}_c = (\v{v}_0+\v{v}_1+\v{v}_2)/3$.

We can rewrite \myrefeq{eq:v3desire} as $ {\v{v}_3 = A\v{v}_0+B\v{v}_1+C\v{v}_2}$, where
\begin{eqnarray}
A&=& I/3 + {(R_{\cEdge{0}{2}} + R_{\cEdge{1}{0}} - 2 R_{\cEdge{2}{1}})}/{9\sqrt{2}}  \nonumber \\
B&=& I/3 + {(R_{\cEdge{1}{0}} + R_{\cEdge{2}{1}} - 2 R_{\cEdge{0}{2}})}/{9\sqrt{2}}  \nonumber \\
C&=& I/3 + {(R_{\cEdge{2}{1}} + R_{\cEdge{0}{2}} - 2 R_{\cEdge{1}{0}})}/{9\sqrt{2}}  \nonumber \ .
\end{eqnarray}
For a new position of $\v{v}_3'$, e.g. later during a simulation run, we
can measure the squared error with 
\begin{equation}
E_{\{\v{v}_3'\}} = \| A\v{v}_0'+B\v{v}_1'+C\v{v}_2' - \v{v}_3' \|^2 .
\end{equation}
Correspondingly, we can compute an overall deformation error for the whole cage 
with $m = (n+1)^3$ new positions $\v{v}'$ with
\begin{equation} \label{eq:defoEne}
E_{defo}(\v{v}') = \frac{1}{n^3}\sum_{i=0}^{n^3} \sum_{j=0}^{8} E_{\{\v{v}_{ij}'\}} /(\frac{1}{n})^2 = \frac{{\v{v}'}^T G \v{v}' }{n} \ .
\end{equation}
For a whole patch cage with $n^3$ cells, we accumulate the deformation energy for the eight corners in each cell.
The energy for a single vertex is given by $E_{\{\v{v}_{ij}'\}}$ above,  where $i$ and $j$ are the index of the cell and its corner respectively.
The right side of \myrefeq{eq:defoEne} is a shortened notation, where $G$ is a $3m\times3m$ matrix containing the accumulated contributions for all points of a cage. 
Since every vertex has at most 6 connected neighbors, every row vector in $G$ has at most 19 entries, corresponding to the x, y and z positions of its neighbors, and a diagonal entry for itself.
Minimizing this quadratic form directly will lead to a trival solution of zero, so it is necessary to solve this problem with suitable constrains.
In practice we want the solution to respect the advected positions. For this we add
an additional advection error $\| \v{v}-\v{v}'\|^2$ that pulls the vertices towards the advected positions, i.e., $\v{v}'$. 
Thus, the total energy we minimize is:
\begin{equation} \label{eq:defoEnergy}
E(\v{v}) = \lambda_0 \frac{{\v{v}}^T G \v{v}}{n} + \frac{1}{m}{\sum \| \v{v}-\v{v}'\|^2} \ ,
\end{equation} 
where $\v{v}'$ is the advected position, $m = (n+1)^3$ is the number of vertices in the grid, and $\lambda_0$ controls the balance 
between advection and deformation. 
Note that in the original formulation, $R_\v{v}$, and thus also $G$, are expressed in terms of $\v{v}$, making the 
whole problem non-linear. Under the assumption that the advected coordinates do not
deform too strongly within a single step, which we found to be a valid assumption even for the large CFL numbers
used in graphics, we linearize the optimization problem by computing $G'$ using $\v{v}'$ in \myrefeq{eq:v3desire}.
The full minimization problem is now given by
\begin{eqnarray}
\frac{\partial E(\v{v})}{\partial \v{v}} = 0 
\ \approx \ 
2[(\lambda G'+I)\v{v} - \v{v}'] , \  \  \lambda = \frac{m}{n} \lambda_0 \ ,
\end{eqnarray}
where we have introduced a scaling factor $\frac{m}{n}$,
that makes the system independent of the chosen cage resolution.
We compute the final position of a patch cage by solving the linear system
$\v{v} = (\lambda G'+I)^{-1} \v{v}'$. Note that this system of equations is relatively small, with $3m$ unknowns per patch.
Hence, it can be solved very efficiently with a few steps of a conjugate gradient solver, independently for all patches.

As our goal is to track large scale motions with our patches, we have to respect
the different spatial scales merged in the flow field. To reduce the effects of small scale
perturbations in the flow, we advect the patches with a low-pass filtered version
of the velocity, where the filter is chosen according to the cage cell size $\Delta x_{\text{cage}}$.

\paragraph{Anticipation} 
To prevent abrupt changes of densities, we fade patches
in and out over a time interval $t_{f}$ for rendering (see below).
\addChanges{For our examples, we use a $t_f$ of $40$ time steps.}
Unfortunately, this temporal fading means that when patches are fully visible, 
they are typically already strongly deformed due 
to the swirling flow motions.
We can  circumvent this problem by letting the patches
{\em anticipate} the flow motion.
I.e., for a new patch at time $t$, we back-track
its motion and deformation to the previous time $t-t_{f}$.
This leads to completely undeformed patch configurations 
exactly at the point in time when they are fully visible.

\paragraph{Initialization}
When seeding a new, undeformed patch at a given position, we found that having 
axis-aligned cages is not the best option. Inspired by classic image feature descriptors, 
we initialize the orientations of our cages according to the gradient of the density field. 
Specifically, we calculate gradient histograms in the cage region $\Omega$, 
and use the top ranked directions as main directions.
\addChanges{The solid angle bins of
3D gradient histograms can be defined using meridians and 
parallels \cite{Scovanner2007SIFT3D}. 
We evenly divide azimuth $\theta_i$ and 
polar angle $\phi_j$ with step sizes $\Delta \theta = \Delta \phi = \pi/n_b$, 
resulting in $2n_b$ and $n_b$ subdivisions for the azimuth and the polar angle, respectively.
3D vectors are then specified as $(r,\theta,\phi)$, where $r$ denotes the radius.
The unit sphere is divided into a set of bins 
$\{ \v{b}_{ij}\}, 0\leq i <2n_b, 0\leq j < n_b$,
where $\v{b}_{ij} = (1, \theta_{i}-\Delta \theta / 2, \phi_{j}-\Delta \phi / 2)$ denotes the 
normalized central direction of the bin in the spherical coordinate system. 
Based on these bins, histograms are calculated as:}
\begin{equation} \label{eq:gradhist}
\resizebox{.9\hsize}{!}{$
h_{ij} = \frac{1}{A_{ij}} \sum_{o \in \Omega} {w\left(\left|\frac{\phi_o - \phi_j}{\Delta \phi} + \frac{1}{2}\right|\right) w\left(\left|\frac{\theta_o - \theta_i}{\Delta \theta} + \frac{1}{2}\right|\right) r_o G\left((\v{x}-\v{o})^2\right)}
$}
\end{equation}
\addChanges{
where $\v{o}$ is the position of a sample point in region $\Omega$ with density 
gradient  $(r_o,\theta_o,\phi_o)$, $ w(d)=\max(0,1-d)$,
$A_{ij}$ represents the solid angle of $\v{b}_{ij}$, and $G$ denotes a Gaussian kernel.}

\addChanges{
At the position $\v{x}$ of a new patch, we compute the gradient histogram for
the smoke density $d_s$ as outlined
above.
}
\addNChanges{
	The histogram has a subdivision of $n_b=16$, and the region $\Omega$ is defined as a $9^3$ grid around $\v{x}$.
	We choose the main direction of the patch as $\v{b}_k$,
	where $k$ denotes the histogram bin with 
	$k = {\arg \max}_{i,j} {\left( h_{ij} \right)}$.}
\addChanges{
For the secondary direction, we recompute the histogram with gradient vectors
in the tangent plane. Thus, instead of $\nabla d_s$ we use 
$(\nabla d_s -(\nabla d_s \cdot \v{b}_x )\v{b}_x)$.
The initial orientation of the patch is then defined in terms of $(\v{b}_x , \v{b}_y, \v{b}_x \times \v{b}_y )$. }
This "gradient-aware" initialization narrows down the potential descriptor space, 
which simplifies the learning problem, and leads to more robust descriptors.

\paragraph{Discussion} 

While as-rigid-as-possible deformations have widely been used for geometric processing task,
our results show that they are also highly useful to track fluid regions while limiting deformation.
In contrast to typical mesh deformation tasks, we do not have handles as constraints, but instead
an additional penalty term that keeps the deformed configuration close to the one pre-scribed
by the flow motion.
The direct comparison between a regular advection, and our cage deformations
is shown in \myreffig{fig:patchCompare}. It is obvious that a direct advection is unusable
in practice, while our deformation limiting successfully lets the cages follow the flow,
while preventing undesirable deformations.

The anticipation step above induces a certain storage overhead,
but we found that it greatly reduces the overall deformation, and hence increase the quality of the synthesized densities. We found the induced memory and storage overheads to be negligible in practice.

\newcommand{\dens}{d}
\newcommand{\wpatch}{w_j}
\newcommand{\wdens}{w_s}
\newcommand{\wrend}{w_r}
\newcommand{\vel}{\vect{u}}

\begin{algorithm}[bt!]
	\SetAlgoNoLine
	\footnotesize 
	Flow quantities: \\
	density grid $\dens$, velocity grid $\vel$, weight grid $\wdens$, patch cages $\v{c}_j$, scalar weights $\wpatch$. \\
	\vspace{5pt}
	{\bf Pre-computation:} \\
	\For{$t=0$ to $t_{\text{end},e}$}{
		Run exemplar simulation, update $\dens_e, \vel_e$ \\
		PatchAdvection($\vel_e$) \\
		SeedNewPatches($\wdens$)\\
		Compute CNN descriptors $\v{d}$ from $\dens_e, \vel_e$ \\
		Save $\v{d}$, and high-resolution $\dens_e$ in patch regions to disk\\
		Accumulate patch spacial weights in ${\wdens}$\\
	}
	\vspace{5pt}
	{\bf Runtime synthesis:} \\
	$G$ = load descriptors from repository\\
	// forward pass: \\
	\For{$t=0$ to $t_{\text{end},t}$}{
		Run source simulation, update $\dens_s, \vel_s$  \\
		PatchAdvection($\vel_s$, $\v{c}$) \\
		SeedNewPatches($\wdens$)\\
		Compute CNN descriptors $\v{d}$ from $\dens_s, \vel_s$ \\
		
		\For{$j=1$ to $\nu_{\text{patches}}$}{
			\If{ Patch $j$ unassigned }{  
				Find closest descriptor to $ \v{d}_j $ in $G$ \\
			} \Else {
				Update and check quality of $ \v{d}_j$ \\
				LimitDeformation($\v{c}_j$) \\
				Update patch fading weights $\wpatch$ (fade out) \\
				Store $j$, $\v{c}_j$, $\wpatch$ for frame $t$ \\
			}
		}
		Store patches at time $t$ , $\dens_s, \vel_s$ \\
		Accumulate patch spacial weights in $\wdens$  \\
	}
	// backward pass: \\
	\For{$t=t_{\text{end},t}$ to $0$}{
		Load $\dens_s, \vel_s$ , patches at time $t$ \\
		\For{$j=1$ to $\nu_{\text{patches}}$}{
			\If{ Anticipation active for Patch $j$ }{  
				PatchAdvection($-\vel_s$, $\v{c}$) \\
				Store $j$, $\v{c}_j$, $\wpatch$ for frame $t$ \\
				Update $\wpatch$ (fade in) \\
		}
		}
	}
	\caption{\addChanges{Pseudo-code for our algorithm.}}
	\label{alg:fullAlg}
	\vspace{\figsp}
\end{algorithm}

\subsection{Synthesis and Rendering}

\addChanges{
In the following, we will outline our two-pass synthesis algorithm, as well as the steps necessary
for generating the final volumes for rendering. A pseudo-code summary of the synthesis step 
is given in \myrefalg{alg:fullAlg}.}

\addChanges{
In the forward pass for flow synthesis, new patches are seeded and advected step by step, and are finally faded out. To seed new patches, a random seeding grid
with spacing 
$s_{\text{p}}/2$ 
is used, $s_{\text{p}} = n \Delta x_{\text{cage}}$ being the size of a patch. In addition, we make use of a patch weighting grid $\wdens$. It uses the native resolution of the simulation, and acts as a threshold to avoid new patches being seeded too closely to existed ones.}
\addChanges{
$\wdens$ accumulates the spatial weights of patches, i.e., spherical kernels centered at the centroids of each patch cage with a radius of $s_{\text{p}}/2$. 
A linear falloff is applied for the last two-thirds of its radius, ramping from one to zero. 
} \addChanges{At simulation time, we typically accumulate the patch weights in $\wdens$ without applying the patch deformations.
This is in contrast to render time, where we deform the patch kernels. The high-resolution weight grid with deformed patch weights for rendering will be denoted $\wrend$ to distinguish it from the low-resolution version $\wdens$.
As $\wdens$ is only used for thresholding the creation of new particles, we found that 
using un-deformed kernels gives very good results with reduced runtimes.
}

\addChanges{
New patches will not be introduced at a sampling position $\v{x_n}$ unless  $\wdens(\v{x}_n)\!<\!0.5$. 
In practice, this means the distance to the closest patch is larger than $s_{\text{p}}/3$. 
For each newly assigned patch, we compute its initial gradient-aligned frame of reference with
\myrefeq{eq:gradhist},
calculate the CNN inputs at this location, and let both CNNs generate the feature descriptors.
Based on the descriptors, we look-up the closest matches from our repository
with a pre-computed kd-tree. For successfully matched patches, 
our deformation limiting advection is performed over their lifetime. 
The maximal lifetimes are determined by the data set lengths of matched repository patches, 
which are typically around 100 frames. 
We remove ill-suited patches, whose re-evaluated descriptor distance is too large for the current flow configuration, or whose deformation energy in \myrefeq{eq:defoEne} exceeds a threshold. In practice, we found 
a threshold of $0.15 s_{\text{p}}^2$ to work well for all our examples.
}
\addChanges{
After the forward pass, matched patches anticipate the motion of the target simulation in a backward pass. Here we move backwards through time, and advect all newly created patches backward over the course of the fade in interval.
Finally, for each frame we store the coarse simulation densities, as well as patch vertex positions $\v{v}$, together with their repository IDs and scalar, temporal fading weights. This data is all that is necessary for the rendering stage.}

\addChanges{
During synthesis the deformation limiting patch advection effectively yields the large motions 
which conform to the input flow, while small scale 
motion is automatically retrieved from the repository frame by frame when we render an image.
Note that we only work with the low-dimensional feature descriptors when synthesizing 
a new simulation result, none of the high-resolution data is required at this stage.}

At render time we synthesize the final high-resolution volume.
To prepare this volume, we also need to consider spatial 
transitions between the patches amongst each other, and
transitions from the patch data to the original simulation. 
For this, we again accumulate the deformed spherical patch kernels into a new grid $\wrend$
with the final rendering resolution, to spatially weight the density data.

\begin{figure}[b]
	\centering 
	\includegraphics[width=0.470 \textwidth]{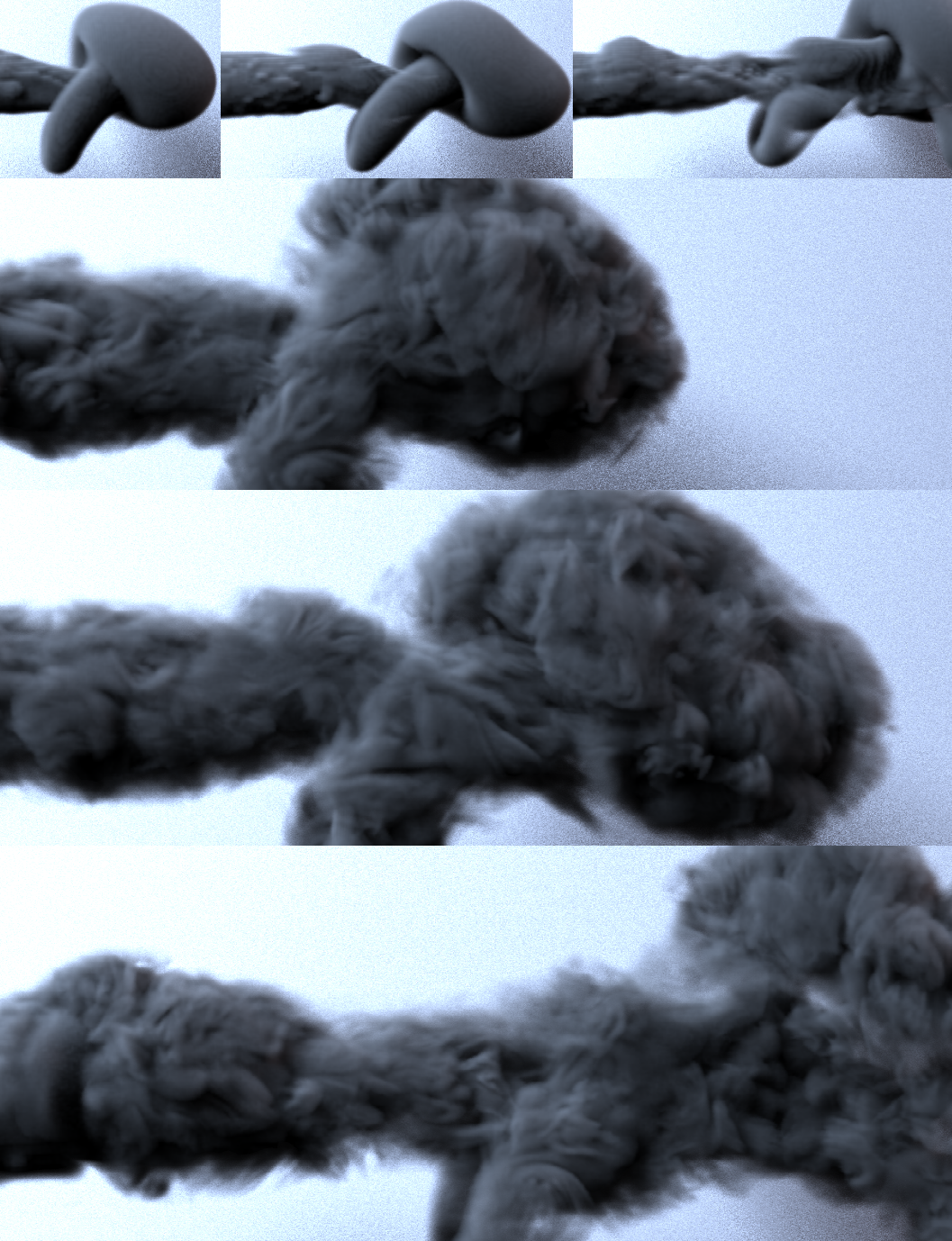}
	\caption{ We apply our method to a horizontal plume simulation. The top
		three images show the input, the bottom three the high-resolution
		densities generated with our method. On average, 388 patches were active 
		per timestep. }
	\label{fig:plume} \vspace{\figsp}
\end{figure}

We then map the accumulate the weighted high-resolution content of a patch data set in the
high resolution rendering volume.
As our repository contains densities that are normalized to $[0..1]$, and our descriptor is invariant to scaling and offset transformations,
we map the repository content to the min-max range of the densities in the target region.
To blend the contributions of overlapping patches, we normalize the accumulated high resolution
density by $\wrend$.
Additionally, we use a blurred version of the original coarse densities
as a mask. We noticed that patches can sometimes drift outside of the main volume while
moving with the flow. The density mask effectively prevents patches from contributing densities 
away from the  original volume.

\section{Results and Discussion}
\label{sec:results}

\begin{table}[b]
	\begin{center}
		\caption{Details of our animation setups and repository data generation.}
		\label{tab:results}
		\resizebox{.92\hsize}{!}{
			\begin{tabular}{ |c|c||c|c|c|c| } 
				\hline
				Scenes    &Fig. & Base res.			& \resizebox{.22\hsize}{!}{base patch res.}	
				& Avg. patches &  \addChanges{Time}\\ \hline
				PlumeX    &\myreffig{fig:plume}& $108\times60\times60$& $15^3$  &   388 & \addChanges{5.3s}\\ 
				Obstacle &\myreffig{fig:obstacle}& $76\times64\times64$ & $12^3$ &	362 & \addChanges{3.9s}\\ 
				Jets &\myreffig{fig:jets}& $90\times60\times60$ & $12^3$ &	486 & \addChanges{4.0s}\\ 
				\hline
				Comp.~$L_2$ &\myreffig{fig:wlt}& $50\times80\times50$ & $9^3$ &	682 & {2.23s}\\
				Comp.~Wlt  &\myreffig{fig:hires}& $50\times80\times50$ & $9^3$ & 647 & {2.42s}\\ 
				\hline
			\end{tabular}
		}
		\resizebox{.92\hsize}{!}{
			\begin{tabular}{ |c||c|c|c|c| } 
				\hline
				  & Res. & Patch res. & Patch no. &  Patch storage \\
				Repo. & $440^3$& $72^3$   & 14894     & 5.1GB densities, 30 MB descriptors\\
				\hline                                                                                   
			\end{tabular}
		}                                                                     
	\end{center}
\end{table}

\begin{figure}[b]
	\centering 
	\includegraphics[width=0.470 \textwidth]{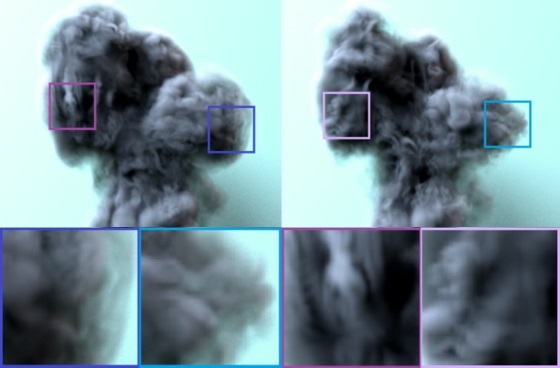}
	\caption{ \addChanges{Using our algorithm with simple descriptors (left) can result
	in overly regular structures (blue rectangles) and sub-optimal matches (purple rectangles). Chances for such patch assignments are reduced with the CNN-based descriptors (right).}}
	\label{fig:l2c}
\end{figure}

\begin{figure}[tb]
	\centering 
	\includegraphics[width=0.470 \textwidth]{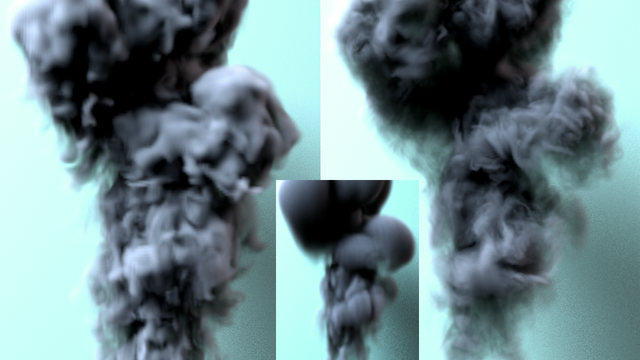}
	\caption{ \addChanges{Based on a simulation with $50\times80\times50$ cells (center), the wavelet turbulence method (left) generates volumes with $150\times240\times150$ cells with 2.75s/frame. Our method (right) yields effective resolutions of $400\times640\times400$ with 2.23s/frame.} }
	\label{fig:wlt}
\end{figure}

\begin{figure*}[tb]
	\centering 
	\includegraphics[width=1.0 \textwidth]{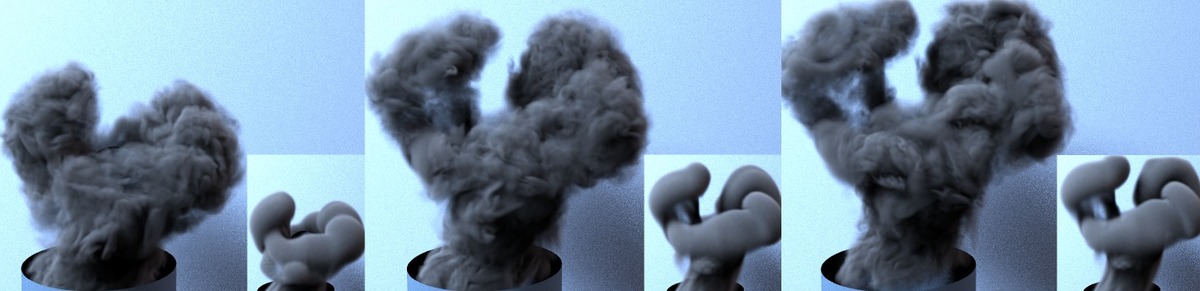}
	\caption{ Our method applied to the flow around a cylindrical obstacle.
		The resolution of the underlying simulation is only  $76\times64\times64$.}
	\label{fig:obstacle} \vspace{\figsp}
\end{figure*}

\begin{figure*}[tb]
	\centering
	\includegraphics[width=1.0 \textwidth]{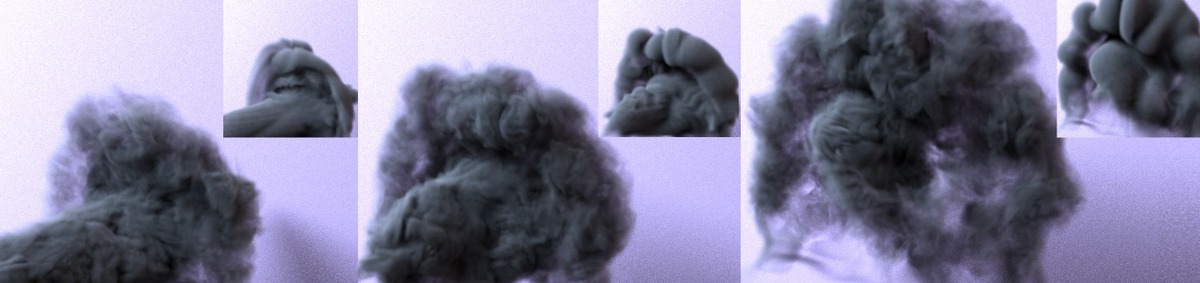}
	\caption{ Two colliding jets of smoke simulated with our approach. The whole simulation including 
		descriptor calculations, look-up, and patch advection took only 4.0 seconds per frame on average. }
	\label{fig:jets} \vspace{\figsp}
\end{figure*}

We will now demonstrate that our approach
can efficiently  synthesize high-resolution volumes for a variety of
different flows. All runtimes in the following are given averages per frame. Typically, we run a single time step per frame of animation.

As a first example we have simulated a simple rising plume example, shown in \myreffig{fig:plume}. Note that gravity acts along the x-axis in our setup.
In this case, the resolution of the original simulation was $108\times60\times60$, and on average, $388$ patches were
active over the course of the simulation. Our approach is able to synthesize a large amount
of clearly-defined detail  to the input flow that does not dissipate over time.  Details of this simulation setup, as well as for the other examples,
can be found in Table~\ref{tab:results}.

A second example is shown in \myreffig{fig:obstacle}. Here we add an additional obstacle,
which diverts the flow. For the patch motion, we simply extend flow velocities into the obstacles,
and add a slight correction along the gradient of the obstacle's signed distance function if a patch
vertex inadvertently ends up inside the obstacle. Our deformation limiting advection smoothly guides the patches
around the obstacle region.

A different flow configuration with colliding jets of smoke is shown in \myreffig{fig:jets}.
For this setup, on average 486 patches were active. 
Note that our patches contain $72^3$ cells in this case. Thus, the effective resolution
for this simulation was ca $560\times360\times360$ cells.
The whole simulation took only 4.0s/frame, which is very efficient given
the high effective resolution. 

\addChanges{For the three examples above, we use $\wdens$ accumulated from deformed patch kernels. By considering the deformation, $\wdens$ better represents the coverage of the patches. However, since we limit deformations, we found that using undeformed kernels generates equivalent visual results with reduced calculation times. Besides, there is still significant room left for accelerating our implementation further.
E.g., we run the fluid solve on the CPU, and we only use GPUs for the CNN descriptor calculations. }

\paragraph{Evaluations:}

\addChanges{
 The CNN descriptors not only increase the recall rate, but also improve the visual quality by retrieving  patches that better adhere to the input flow. This can be seen in \myreffig{fig:l2c}, where our result
 is shown on the right, while the left hand side
 uses our full pipeline with a simplified distance calculation, i.e., without the use of a CNN.
Instead, we use an $L_2$ distance of down-sampled versions of $\dens_s$ and the curl of $\vel_s$. These values are normalized and used as descriptors directly. Specifically, we down-sampled $\dens_s$ to a resolution of $7^3$, and the curl to $5^3$, resulting in a combined descriptor with $(343+375)$ entries.
This is similar to the size of our CNN-based descriptors.
Since CNN descriptors have a good understanding of correlation between different fluid resolutions, they offer results with small-scale vortices and vivid structures that fit the target well, while the simple descriptors sometimes offer plain and noisy structures (the blue regions in \myreffig{fig:l2c}). 
Additionally, the simple descriptors can introduce un-plausible motions, which becomes apparent
in the regions marked in purple in \myreffig{fig:l2c}. There, we know from theory
that the baroclinic contribution to vorticity should be along the cross product of density
and pressure gradient. 
Thus, the vertical structures caused by the simple descriptors are not plausible for the buoyancy
driven input simulation.}

\addChanges{Based on the setup from \myreffig{fig:l2c},
 we further compare our approach to the wavelet turbulence me\-thod~\cite{Kim:2008:wlt} as a representative from the class of up-res methods.
In order to make the approaches comparable, we consider their performance given a limited 
and equivalent computational budget. 
With this budget the wavelet turbulence method produces results with a 3 times higher resolution, consuming 2.75s/frame. This is close to the 2.23s/frame our method requires. However, our method effectively yields an 8 times higher resolution (see \myreftab{tab:results}).
For our simulation, this setup used the $\wdens$ field with undeformed kernel evaluations.
 Apart from the difference in detail, 
the wavelet turbulence version exhibits a noticeable
deviation from the input flow in the lower part of the volume. 
Here numerical diffusion accumulates to cause significant drift over time, 
while our method continues to closely conform to the input.}

\begin{figure}[tb]
	\centering 
	\includegraphics[width=0.470 \textwidth]{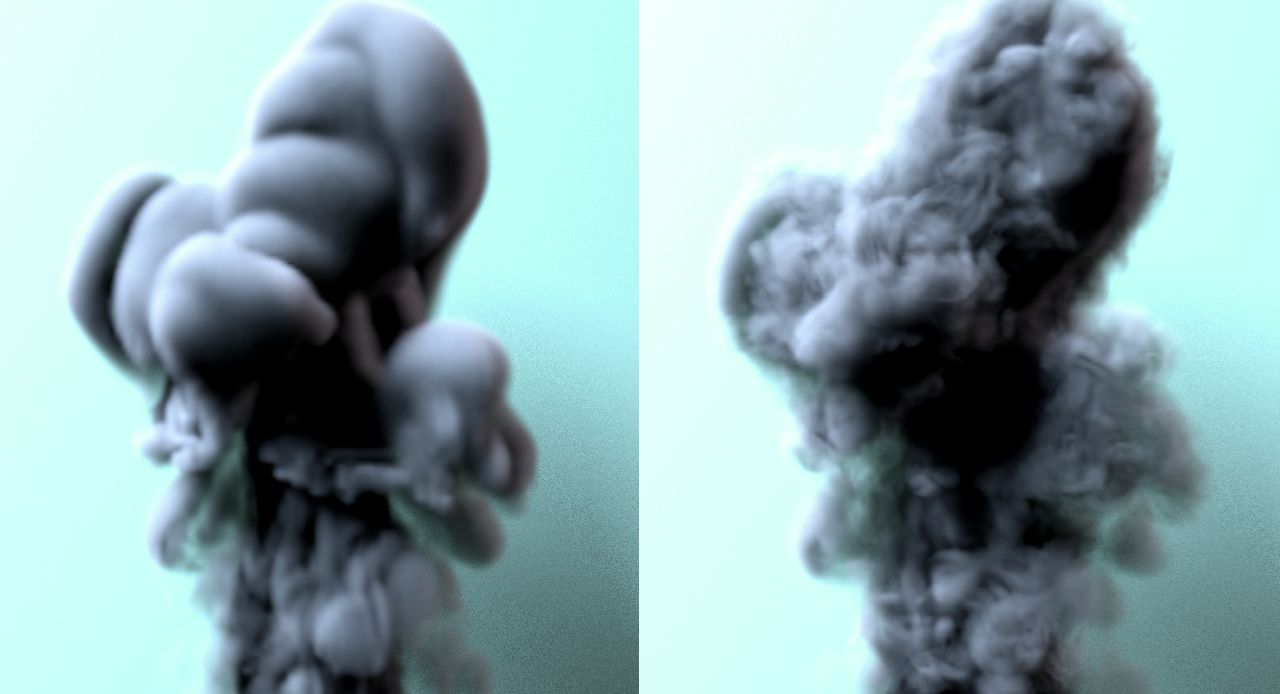}
	\caption{ \addChanges{In order to compare  a full simulation of $100\times160\times100$  (left) with our approach, we downsample the full simulation to $50\times80\times50$, and then apply our algorithm (right). The latter spends 2.42s/frame, while the full simulation requires 2.51s/frame.} }
	\label{fig:hires}
\end{figure}

\addChanges{
Finally, we compare our method to a regular simulation 
with doubled resolution. As expected, this version results in a 
 different large scale motion, and  
 in order to compare the outputs, we 
applied our CNN based synthesis method
to a down-sampled version of the high resolution simulation. 
While the regular high resolution scene spends 2.5s/frame, our method takes only 2.42s, but offers fine details, as shown in \myreffig{fig:hires}. }
\paragraph{Limitations and Discussion:}
\addChanges{One limitation of our approach is that we cannot guarantee fully divergence-free motions on small scales. For larger scales, our outputs conform to the original, diver\-gence-free motion. The small scale motions contained in repository patches are likewise recorded from fully divergence-free flows, but as our patches deform slightly, the resulting motions are not guaranteed to be divergence-free. Additionally, spatial blending can introduce regions with divergent motions. 
Our algorithm shares this behavior with other synthesis approaches, e.g., texture synthesis. 
However, as we do not need to compute an advection step based on these motions, our method avoids accmuluating mass losses (or gains) over time.
}

There are many avenues for smaller improvements of our neural network approach, e.g., applying techniques such as batch normalization, or specialized techniques
for constructing the training set. 
However, we believe that our current approach
already demonstrates that deforming Lagrangian patch regions 
are an excellent basis for CNNs in conjunction with fluid flows.
It effectively makes our learning approach invariant to large scale motions. Removing
these invariants for machine learning problems is an important topic, as mentioned e.g.
by Ladicky et al. \shortcite{ladicky2015data}. 
Apart from motion invariance, we arrive at an algorithm
that can easily applied to any source resolution. For other CNN-based
approaches this is typically very difficult to achieve, as networks are specifically trained
for a fixed input and output size \cite{tompson2016accelerating}.

\addChanges{ Due to its data-driven nature, our method requires more hard-disk space than procedural methods. As shown in \myreftab{tab:results}, the density data of the patches in our 3D repository takes up ca. 5.1GB of disk space. Fortunately, we only load the descriptors at simulation time: ca. 15MB for density descriptors, and another 15MB for curl descriptors. At render time, we have to load ca. 400 data sets per frame, i.e. ca. 137MB in total. }

Another consequence of our machine-learning approach is that
our network is specifically adapted to a set
of algorithmic choices. Thus, for a different solver, 
it will be advantageous to re-train the network with a suitable data set and adapted
interval $t_r$. It will be an interesting area of future work whether a sufficiently deep CNN
can learn to compute descriptors for larger classes of solvers.

\section{Conclusions}
\label{sec:conclusions}

We have presented a novel CNN-based method to realize a  
practical data-driven workflow 
for smoke simulations. Our approach is the first one to use
a flow repository of space-time datasets to synthesize high-resolution results 
with only a few seconds of runtime per frame.
At the same time, our work represents a first demonstration
of the usefulness of descriptor learning in the context of fluids flows, 
and we have shown that it lets us establish 
correspondences between different simulations in the presence of numerical viscosity.
As our approach is a data-driven one, it can be used for any choice of Navier-Stokes solver, as long
as enough input data is available. Additionally, the localized descriptors make our 
approach independent of the simulation resolution.

We believe the direction of data-driven flow synthesis is a very promising area
for computer graphics applications. Art-directable solvers that are at the same
time fast and stable are in high demand. In the future, we believe it will be very interesting
to extend our ideas towards stylization of flows, and towards synthesizing not only
an advected quantity such as smoke density, but the flow velocity itself.

\begin{acks}

The authors would like to thank Marie-Lena Eckert for helping with generating the video, and Daniel Hook for his work on the direct density synthesis with CNNs.

\end{acks}

\bibliography{CG_and_CV_conferences_and_journals,fluids}

\end{document}